\documentclass{jfm}
\usepackage{graphicx}
\usepackage{epstopdf, epsfig}
\usepackage{multirow}
\usepackage{mathrsfs,amsmath}
\usepackage{bm}
\usepackage{amssymb}
\usepackage{multirow,bigdelim}
\usepackage{upgreek}
\usepackage{units}
\usepackage{array,multirow}
\usepackage{booktabs}
\usepackage{psfrag}
\usepackage{tikz}
\usetikzlibrary{arrows.meta}
\usepackage{color}
\usepackage{xcolor}
\usepackage{threeparttable}	
\usepackage{dcolumn}		
\newcolumntype{d}[1]{D{.}{.}{#1}}
\usetikzlibrary{decorations}
\usetikzlibrary{decorations.markings}
\usetikzlibrary{calc,decorations.pathreplacing}
\usepackage{lipsum}
\usepackage{mathtools}
\usepackage{ar}
\usepackage{enumitem}
\usepackage{url}
\usepackage{wasysym}

\definecolor{Bblack}{rgb}{0.00, 0.00, 0.00}
\definecolor{Bblue}{rgb}{0,0.572,0.812}
\definecolor{Bgrey}{rgb}{0.25, 0.25, 0.25}

\newcommand*{\myfont}{\fontfamily{ptmri}\selectfont}
\DeclareTextFontCommand{\textmyfont}{\myfont}

\shorttitle{Decomposition of the streamwise TKE in boundary layers. Part 2}
\shortauthor{W. J. Baars and I. Marusic}
\title{Data-driven decomposition of the streamwise turbulence kinetic energy in boundary layers. Part 2. Integrated energy and $A_1$}

\author{Woutijn J. Baars$^1$
  \corresp{\email{baars@eng.au.dk}}
  \and
  Ivan Marusic$^2$}

\affiliation{$^1$Department of Engineering, Aarhus University, 8000 Aarhus C, Denmark \\[\affilskip] $^2$Department of Mechanical Engineering, The University of Melbourne, VIC 3010, Australia}

\begin{document}

\maketitle
\begin{abstract}
Scalings of the streamwise velocity energy spectra in turbulent boundary layers were considered in Part~1. A spectral decomposition analysis provided a means to separate out attached and non-attached eddy contributions and was used to generate three spectral sub-components, one of which is a close representation of the spectral signature induced by self-similar, wall-attached turbulence. Since sub-components of the streamwise turbulence intensity $\overline{u^2}$ follow from an integration of the velocity energy spectra, we here focus on the scaling of the former. Wall-normal profiles and Reynolds number trends of the three individual, additive sub-components of the streamwise turbulence intensity are examined. This allows for revisiting the scaling of the turbulence intensity in more depth, in comparison to evaluating the total streamwise turbulence intensity. Based on universal trends across all Reynolds numbers considered, some evidence is given for a Townsend--Perry constant of $A_1 = 0.98$, which would describe the wall-normal logarithmic decay of the turbulence intensity per Townsend's attached-eddy hypothesis. It is also demonstrated how this constant can be consistent with the Reynolds-number increase of the streamwise turbulence intensity in the near-wall region.
\end{abstract}

\section{Introduction and context}\label{sec:intro}
Wall-normal trends of the streamwise turbulence intensity (TI), denoted as $\overline{u^2}$, are a prerequisite to modelling efforts of wall-bounded turbulence. Several models for $\overline{u^2}$ are hypothesis-based. For instance, the model of \citet{marusic:2003a} was inspired by the attached-eddy hypothesis \citep[AEH,][]{townsend:1976bk}, while \citet{monkewitz:2015a} constructed a model via asymptotic expansions and \citet{chen:2018a} via a dilation symmetry approach. The works of \citet{vassilicos:2015a} and \citet{laval:2017a} derived a model for the streamwise TI by introducing a new spectral scaling at the very large-scale end of the spectrum, beyond the scales associated with a $k_x^{-1}$ region. All models require validation and calibration for the streamwise TI \citep{monkewitz:2017c} and assumptions are inevitable for extrapolated conditions. More importantly, validation of the underlying spectra are often avoided, which could result in questioning of the model assumptions. Even with available wall-normal profiles of $\overline{u^2}$ and its spectral distribution, from both numerical computations and experiments \citep[\emph{e.g.}][]{marusic:2010a}, definitive scalings remain elusive and continue to be of research interest. The difficulty in finding empirical scaling trends is mainly due to the weak dependence of $\overline{u^2}$ on the Reynolds number, the limited Reynolds-number range over which direct numerical simulations are feasible/available and the practical challenges associated with experimental acquisition of fully-resolved data. 

Velocity energy spectra inform how the streamwise TI is distributed across wavenumbers, because the streamwise TI (the velocity variance or normal stress) equates to the integrated spectral energy via Parselval's theorem (\emph{e.g.} $\overline{u^2} = \int \phi_{uu} {\rm d}k_x$). Part~1 considered the streamwise velocity energy spectra---and in particular by way of a spectral decomposition to separate out several wall-attached and non-attached eddy contributions---thus allowing an evaluation of the wall-normal profiles and Reynolds number trends of three individual, additive sub-components of the streamwise TI.

First, this introduction addresses the widely researched logarithmic decay of the streamwise TI within the outer region of turbulent boundary layers (TBLs) in \S\,\ref{sec:A1issue}. Then, we discuss the contentious issue of the $k_x^{-1}$ scaling in the streamwise velocity spectra $\phi_{uu}(k_x)$, where $k_x$ is the streamwise wavenumber (and $\lambda_x \equiv 2\pi/k_x$ is the streamwise wavelength). We briefly review Part~1 \citep{baars:part1} in \S\,\ref{sec:trdecom}, which presented a data-driven spectral decomposition.

Notation in this paper is identical to that used in Part~1. Coordinates $x$, $y$ and $z$ denote the streamwise, spanwise and wall-normal directions of the flow, whereas the friction Reynolds number $Re_\tau \equiv \delta U_\tau/\nu$ is the ratio of $\delta$ (the boundary layer thickness) to the viscous length-scale $\nu/U_\tau$. Here $\nu$ is the kinematic viscosity and $U_\tau = \sqrt{\tau_o/\rho}$ is the friction velocity, with $\tau_0$ and $\rho$ being the wall-shear stress and fluid's density, respectively. When a dimension of length is presented in \emph{outer-scaling}, it is normalized with scale $\delta$, while a \emph{viscous-scaling} with $\nu/U_\tau$ is signified with superscript `+'. Lower-case $u$ represents the Reynolds decomposed fluctuations and $U$ the absolute mean.

\subsection{Townsend--Perry constant $A_1$ in the context of the turbulence intensity and spectra}\label{sec:A1issue}
\citet{townsend:1976bk} hypothesized that the energy-containing motions in TBLs are comprised of a hierarchy of geometrically self-similar eddying motions, that are inertially dominated (inviscid), attached to the wall and scalable with their distance to the wall \citep{marusic:2019a}. According to the classical model of attached eddies \citep{perry:1982a}, the wall-normal extents of the smallest attached eddies scale with inner variables, \emph{e.g.}, $100\nu/U_\tau$, while the largest scale on $\delta$. Consequently, $Re_\tau$ is a direct measure of the attached-eddy range of scales. Following the attached-eddy modelling framework, the streamwise TI within the logarithmic region adheres to 
\begin{eqnarray}
 \label{eq:u2log}
 \overline{u^2}^+ = B_1 - A_1\ln\left(\frac{z}{\delta}\right),
\end{eqnarray}
where $A_1$ and $B_1$ are constants; $A_1$ was dubbed the Townsend--Perry constant. A scaling of $\phi_{uu} \propto k_x^{-1}$ (or a plateau in the premultiplied spectrum $k^+_x\phi^+_{uu}$) is consistent with the presence of a sufficient range of attached-eddy scales. Such a spectral scaling for the energy-containing, inertial range of anisotropic scales can be predicted with the aid of dimensional analysis, a spectral overlap argument and an assumed type of eddy similarity \citep[\emph{e.g.}][]{perry:1975a,davidson:2009a}. \citet{perry:1986a} related the plateau magnitude of the premultiplied spectrum back to (\ref{eq:u2log}), resulting in
\begin{eqnarray}
 \label{eq:kmin1}
 k^+_x\phi^+_{uu} = A_1.
\end{eqnarray}
An underlying assumption of (\ref{eq:u2log}), in combination with (\ref{eq:kmin1}), is that all energy is induced by self-similar attached-eddy motions. And so, from detailed studies on the streamwise turbulence kinetic energy, from which profiles of $\overline{u^2}(z)$ and streamwise spectra $\phi_{uu}$ are available, the Townsend--Perry constant $A_1$ inferred via either (\ref{eq:u2log}) or (\ref{eq:kmin1}) must be equal, of course provided that attached-eddy turbulence dictates the scaling.
\begin{figure} 
\vspace{10pt}
\centering
\includegraphics[width = 0.999\textwidth]{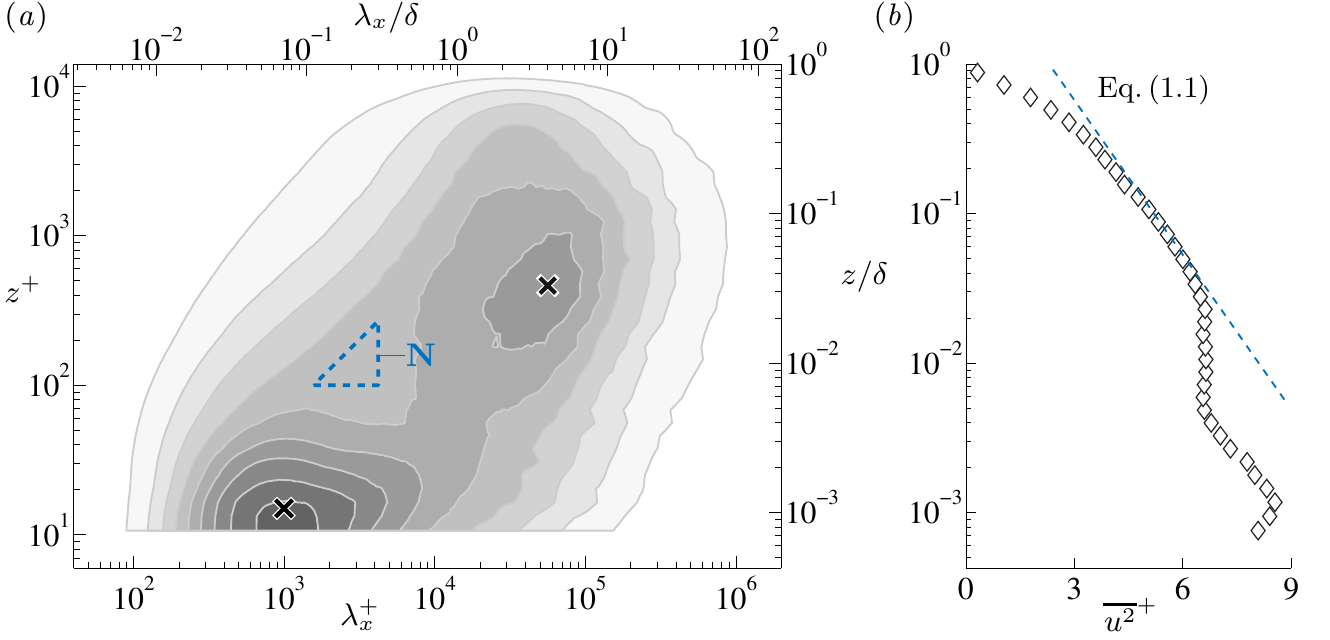}
   \caption{(\emph{a}) Premultiplied energy spectrogram $k^+_x\phi^+_{uu}$ (filled iso-contours 0.2:0.2:1.8) at $Re_\tau \approx 14\,100$ \citep{baars:2017a}. Triangle `N' refers to the $k_x^{-1}$ region identified by \citet{nickels:2005a}. (\emph{b}) Turbulence intensity profile and (\ref{eq:u2log}) with $A_1 = 1.26$ and $B_1 = 2.30$.}
   \label{fig:spintro1}
\end{figure}

Thus far, evidence for (\ref{eq:kmin1}) has been inconclusive, mainly due to the limited spectral range over which this region may exist. It is instructive to present an energy spectrogram: premultiplied spectra at 40 logarithmically-spaced positions within the range $10.6 \apprle z^+ \apprle \delta^+$ are presented with iso-contours of $k^+_x\phi^+_{uu}$ in Figure~\ref{fig:spintro1}(\emph{a}). These spectra were obtained from hot-wire measurements at $Re_\tau \approx 14\,100$ in Melbourne's TBL facility \citep{baars:2017a}. Near-wall streaks \citep{kline:1967a} dominate the inner-spectral peak in the TBL spectrogram (identified with the $\times$ marker at $\lambda_x^+ = 10^3$ and $z^+ = 15$), while large-scale organized motions induce a broad spectral peak in the log-region, indicated with a $\times$ marker at $\lambda_x = 4\delta$ and $z^+ = 3.9 Re_\tau^{\nicefrac{1}{2}} \approx 464$ \citep{mathis:2009a}. \citet{nickels:2005a} determined a $k_x^{-1}$ region as $z^+ > 100$, $\lambda_x > 15.7z$ (wall-scaling) and $\lambda_x < 0.3\delta$ (outer-scaling) at $Re_\tau \approx 14\,000$ (triangular region `N' in Figure~\ref{fig:spintro1}\emph{a}). This region satisfied (\ref{eq:kmin1}) with $k^+_x\phi^+_{uu} = A_1 \approx 0.92$. In Part~1 it was determined from a coherence analysis \citep{baars:2017a2}, relative to a wall-based reference, that wall-attached self similar motions only become spectrally energetic at $\lambda_x \apprge 14z$. It is important to note here that this result was derived from a stochastic analysis of the streamwise velocity component only. It was furthermore suggested that $\phi_{uu} \propto k_x^{-1}$ is unlikely for $Re_\tau \apprle 80\,000$. This is consistent with the study of \citet{chandran:2017a}, where experimentally acquired streamwise--spanwise 2D spectra of $u$ at $Re_\tau \sim \mathcal{O}(10^4)$ were examined for a $k_x^{-1}$. They concluded that an appreciable $k^{-1}$ scaling region can only appear for $Re_\tau \apprge 60\,000$. Moreover, even for the highest $Re_\tau$ laboratory data, the presence of a $\phi_{uu} \propto k_x^{-1}$ has been inconclusive \citep{morrison:2002a,rosenberg:2013a,vallikivi:2015a}, while this region must grow with $Re_\tau$.

We now switch our attention to evidence for (\ref{eq:u2log}). A caveat in determining $A_1$ from $\overline{u^2}(z)$ profiles is that, generally, all turbulent scales are lumped together as one (integral of the entire spectrum). This approach inherently assumes that the attached-eddy structures dominate $\overline{u^2}$. Now, if one accepts that the attached-eddy contribution to the overall turbulence intensity grows with $Re_\tau$, this assumption should become more valid. For this reason, \citet{marusic:2013a} considered high $Re_\tau$ data in the range $2 \times 10^4 < Re_\tau < 6 \times 10^5$ \citep{winkel:2012a,hultmark:2012a,hutchins:2012a,marusic:2015a} and inferred that $A_1 = 1.26$ (see Figure~\ref{fig:spintro1}\emph{b}). It is worth noting that the value for $A_1$ has changed significantly over time. For instance, values for $A_1$ have been quoted as 1.03 \citep{perry:1990a}, 1.26 \citep{hultmark:2012a,marusic:2013a,orlu:2017a} and 1.65 \citep{yamamoto:2018a}. These variations in $A_1$ are largely due to the varying TI slope with $Re_\tau$ and the different fitting regions for (\ref{eq:u2log}).

The previous discussion illustrates that $A_1$ values found from $\overline{u^2}$ profiles vary, while the AEH envisions a constant $A_1$ in (\ref{eq:u2log}): one that is invariant with $Re_\tau$. Moreover, $A_1$ values found from $\overline{u^2}$ profiles do not agree with values for $A_1$ inferred from spectra via (\ref{eq:kmin1}), despite that this is expected per the attached-eddy model \citep{perry:1986a}. A central facet of this mismatch is the simple fact that (\ref{eq:u2log}) and (\ref{eq:kmin1}) are restricted to attached-eddy turbulence only, while in measures of the total streamwise turbulence kinetic energy more contributions are present. For a quantitative insight into what portion of the turbulence kinetic energy is representative of attached-eddy turbulence, a spectral decomposition method was introduced in Part~1 (and is summarized next).

\subsection{Streamwise energy spectra and the triple decomposition}\label{sec:trdecom}
Data-driven spectral filters were empirically found with the aid of two-point measurements and a spectral coherence analysis. A first filter, denoted as $f_\mathcal{W}$, was based on a reference position deep within the near-wall region (or at the wall). Such a reference position allows for determining the degree of coherence between the $u$ fluctuations within the TBL and the fluctuations that are present at the reference position. The other filter, $f_\mathcal{L}$, was based on a reference position in the logarithmic region. It was verified that both spectral filters were universal for $Re_\tau \sim \mathcal{O}(10^3) - \mathcal{O}(10^6)$. Filter $f_\mathcal{W}$ was formulated as
\begin{eqnarray}
\label{eq:Wfilt1}
f^p_{\mathcal{W}}\left(z;\lambda_x\right) = 
     \begin{cases}
       0 & \quad \lambda_x < Rz \\
       {\rm min}\left\lbrace C_1\ln\left(\frac{\lambda_x}{z}\frac{1}{R}\right),\,1\right\rbrace & \quad Rz \leq \lambda_x \leq T_n\delta \\
       {\rm min}\left\lbrace C_1\ln\left(\frac{T_n\delta}{z}\frac{1}{R}\right),\,1\right\rbrace & \quad \lambda_x > T_n\delta \\
     \end{cases}
\end{eqnarray}
Subscript $\mathcal{W}$ signifies the wall-based reference, on which this filter is based, and the three constants are: $C_1 = 0.3017$, $R = 14.01$ and $T_n = 10$ (Table~1, Part~1). A smooth filter $f_{\mathcal{W}}\left(z;\lambda_x\right)$ was generated by convoluting (\ref{eq:Wfilt1}) with a log-normal distribution, $g(\lambda_x)$, spanning six standard deviations, corresponding to 1.2 decades in $\lambda_x$ (details are provided in Part~1). Filter $f_{\mathcal{W}}\left(z;\lambda_x\right) \in [0,1]$ and equals a wavelength-dependent fraction of energy that is stochastically \emph{coherent} with the near-wall region. Consequently, $(1-f_{\mathcal{W}})$ is the \emph{incoherent} energy fraction. Filter $f_\mathcal{L}$ employs a reference position $z_\mathcal{L}$ in the logarithmic region:
\begin{eqnarray}
\label{eq:Lfilt1}
f^p_{\mathcal{L}}\left(z_{\mathcal{L}};\lambda_x\right) = 
     \begin{cases}
       0 & \quad \lambda_x < R'z_{\mathcal{L}} \\
       {\rm min}\left\lbrace C'_1\ln\left(\frac{\lambda_x}{z_{\mathcal{L}}}\frac{1}{R'}\right),\,1\right\rbrace & \quad R'z_{\mathcal{L}}\leq \lambda_x \leq T_n\delta \\
       {\rm min}\left\lbrace C'_1\ln\left(\frac{T_n\delta}{z_{\mathcal{L}}}\frac{1}{R'}\right),\,1\right\rbrace  & \quad \lambda_x > T_n\delta \\
     \end{cases}
\end{eqnarray}
Filter constants are $C'_1 = 0.3831$, $R' = 13.18$ and $T_n = 10$. A smooth filter $f_{\mathcal{L}}\left(z_{\mathcal{L}};\lambda_x\right)$ was formed in a similar way as $f_{\mathcal{W}}\left(z;\lambda_x\right)$. Of the fraction of energy that is stochastically coherent with the near-wall region (via $f_\mathcal{W}$), a sub-fraction of that energy is also coherent with $z_{\mathcal{L}}$ in the logarithmic region (and this fraction is prescribed by $f_\mathcal{L}$). 

\begin{figure} 
\vspace{10pt}
\centering
\includegraphics[width = 0.999\textwidth]{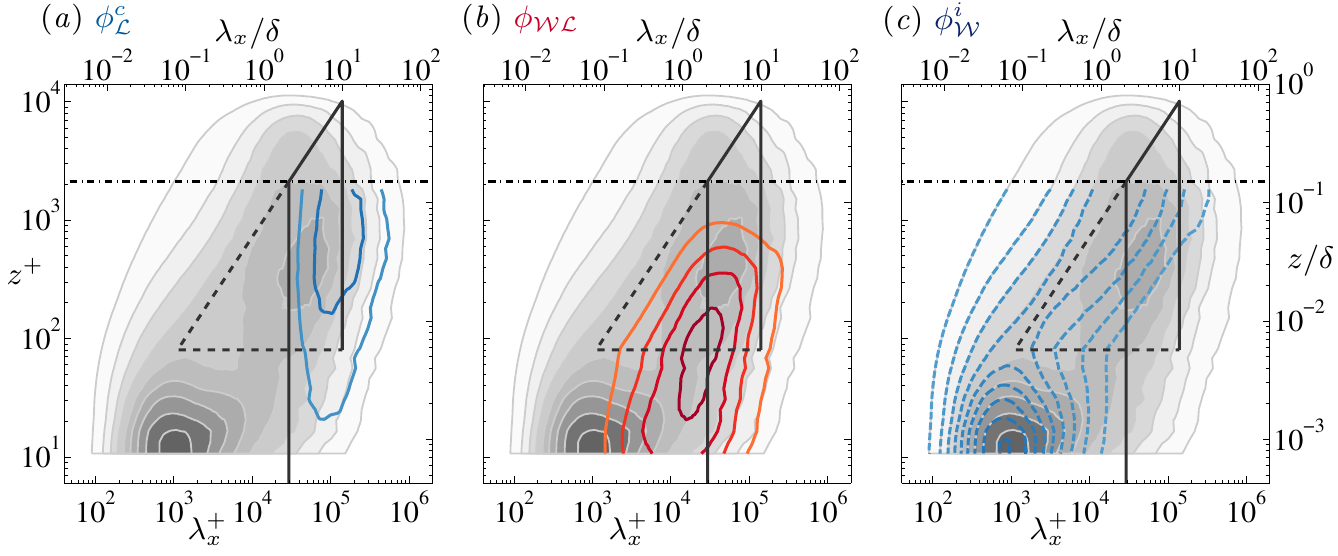}
   \caption{Dataset $\mathcal{W}$ with $Re_\tau \approx 14\,100$. (a-c) Premultiplied energy spectrograms of the three spectral sub-components (for $z < z_{\mathcal{L}}$), each of them overlaid on the total energy spectrogram (filled iso-contours 0.2:0.2:1.8). Following Figure~14 of Part 1.} 
   \label{fig:trdecom}
\end{figure}

A triple decomposition for $\phi_{uu}$ was formed from $f_\mathcal{W}$ and $f_\mathcal{L}$ (\S\,5.1, Part~1), following
\begin{eqnarray}
 \label{eq:trdecom1}
 \phi_{\mathcal{L}}^c\left(z;\lambda_x\right) \equiv &\phi_{uu}\left(z;\lambda_x\right)&f_{\mathcal{L}}\left(z;\lambda_x\right) \\
 \label{eq:trdecom2}
 \phi_{\mathcal{W}}^i\left(z;\lambda_x\right) \equiv &\phi_{uu}\left(z;\lambda_x\right)&\left[1-f_{\mathcal{W}}\left(z;\lambda_x\right)\right] \\
 \label{eq:trdecom3}
 \phi_{\mathcal{W}\mathcal{L}}\left(z;\lambda_x\right) \equiv &\phi_{uu}\left(z;\lambda_x\right)&\left[f_{\mathcal{W}}\left(z;\lambda_x\right) - f_{\mathcal{L}}\left(z;\lambda_x\right)\right]
\end{eqnarray}
Consequently, $\phi_{uu} = \phi_{\mathcal{L}}^c + \phi_{\mathcal{W}\mathcal{L}} + \phi_{\mathcal{W}}^i$ and Figure~\ref{fig:trdecom} illustrates this decomposition for $Re_\tau \approx 14\,100$ (duplicate of Figure~14, Part~1). The three energy spectrograms of (\ref{eq:trdecom1})--(\ref{eq:trdecom3}) are overlaid on the premultiplied energy spectrogram $k^+_x\phi^+_{uu}$. Here, $z_\mathcal{L} = 0.15\delta$ and the triple-decomposition is performed for $z < z_\mathcal{L}$. In the near-wall region, here taken as $z^+ \apprle z^+_T$ (nominally $z^+_T = 80$ is used, roughly the wall-normal position at which the near-wall spectral peak becomes indistinguishable from the spectrogram), $f_{\mathcal{W}}$ is $z$-invariant and taken as $f_{\mathcal{W}}(z^+_T;\lambda_x)$. Throughout this work, the exact value of $z^+_T$ is of secondary importance, since small variations in this location do not affect conclusions, given a lower bound of the logarithmic region in viscous scaling, $z_T = \mathcal{O}(100\nu/U_\tau)$. In \S\,2.2 of Part~1 we discussed that a classical scaling of the lower limit of the logarithmic region (as opposed to a meso layer type scaling via $z^+ \propto Re_\tau^{0.5}$) should not be discarded. In fact, in this paper we show that when we accept such a classical scaling, the growth of the near-wall peak in $\overline{u^2}$ can be explained via the increasingly intense energetic imprint of the Reynolds number dependent outer motions onto the near-wall region.

Component $\phi_{\mathcal{L}}^c$ (Figure~\ref{fig:trdecom}\emph{a}) comprises the energy that is coherent via $f_{\mathcal{L}}$: large-scale \emph{wall-attached} energy that is coherent with $z_{\mathcal{L}} = 0.15\delta$. This component includes spectral imprints of self-similar, wall-attached structures reaching beyond $z_{\mathcal{L}}$ and non-self-similar wall-attached structures that are coherent with $z_{\mathcal{L}}$ (\emph{e.g.} some VLSMs). Component $\phi_{\mathcal{W}}^i$ (Figure~\ref{fig:trdecom}\emph{c}) is formed from the $f_{\mathcal{W}}$-based incoherent energy. This small-scale energy is \emph{wall-detached} and includes detached (non)-self-similar motions, such as phase-inconsistent attached eddies, incoherent VLSMs, etc. The remaining component, $\phi_{\mathcal{W}\mathcal{L}}$, equals the wall-coherent energy below $z_{\mathcal{L}}$ and consists of self-similar and non-self-similar contributions. However, the non-self-similar contributions are likely to reside at large $\lambda_x$ \citep[reflecting global modes,][]{bullock:1978a,delalamo:2003a}. 

\subsection{Present contribution and outline}\label{sec:outline}
Coming back to \S\,\ref{sec:A1issue}, we can now argue that $A_1$ can be inferred from $\overline{u^2}(z)$ profiles via (\ref{eq:u2log}), as long as the streamwise TI contributions, other than the one from the self-similar wall-attached motions, are removed. This step is crucial, because Part~1 already addressed that the other contributions (\emph{e.g.} $\phi^c_\mathcal{L}$ in Figure~\ref{fig:trdecom}\emph{a} and $\phi^i_\mathcal{W}$ in Figure~\ref{fig:trdecom}\emph{c}) result in additions to the streamwise TI that constitute a clear $z$-dependence. And, the Reynolds-number dependent outer-spectral peak seems to mask a possible $\phi_{uu} \propto k_x^{-1}$ \citep[see spectra in][]{morrison:2002a,nickels:2005a,marusic:2010a2,baidya:2017a,samie:2018a}. When re-assessing $A_1$ in this paper, both the spectral view and $\overline{u^2}(z)$ are considered simultaneously, while recognizing that $A_1$ must solely be associated with the turbulence that obeys the AEH.

Next, in \S\S\,\ref{sec:scaling}-\ref{sec:scalingRe}, decompositions of the streamwise TI are presented for a range of $Re_\tau$. Data used are the same as in Part 1 \citep[][\S\,3.2]{baars:part1}. Findings on the Townsend--Perry constant $A_1$ are reconciled in \S\,\ref{sec:A1summ}, after which its relation to the near-wall TI growth, with $Re_\tau$, is presented in \S\,\ref{sec:nearwall}. Empricial trends within the wall-normal profiles for all three additive sub-components of the streamwise TI are presented in \S\,\ref{sec:empmodel}, together with a discussion of their scalings.

\section{Decomposition of the streamwise turbulence intensity}\label{sec:scalingO}
\subsection{Methodology and logarithmic scalings}\label{sec:scaling}
\begin{figure} 
\vspace{10pt}
\centering
\includegraphics[width = 0.999\textwidth]{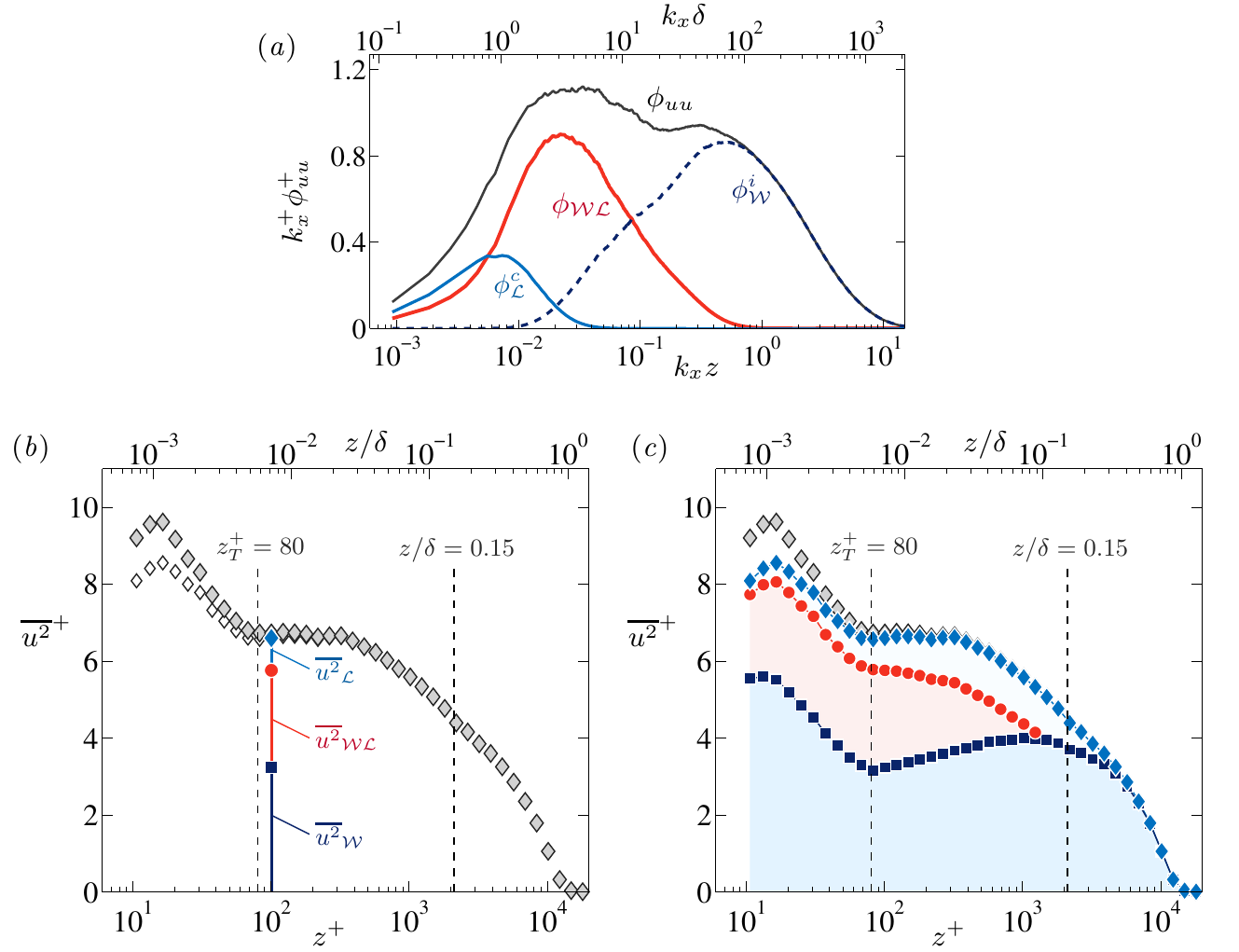}
   \caption{(\emph{a}) Triple-decomposed energy spectrum at $z^+ \approx 101$ and $Re_\tau \approx 14\,100$, reproduced from Figure~14(\emph{e}) in Part 1. (\emph{b}) Streamwise TI profile with the three TI sub-components following (\ref{eq:u2decom2}). (\emph{c}) Similar to (\emph{b}) but for all wall-normal locations.}
   \label{fig:TKEdecom}
\end{figure}

Figure~\ref{fig:TKEdecom}(\emph{a}) shows the three sub-components $\phi^c_\mathcal{L}$, $\phi_{\mathcal{W}\mathcal{L}}$ and $\phi^i_\mathcal{W}$ for the spectrum at $z^+ \approx 101$ (slice through Figure~\ref{fig:trdecom}). When integrated, these sub-components form three contributions to the streamwise TI, being $\overline{u^2}_{\mathcal{L}}$, $\overline{u^2}_{\mathcal{W}\mathcal{L}}$ and $\overline{u^2}_{\mathcal{W}}$, respectively. In summary:
\begin{eqnarray}
 \label{eq:u2decom1}
 \overline{u^2}\left(z\right) &=& \int \phi_{uu}\left(z;k_x\right) {\rm d}k_x \\ \label{eq:u2decom2} &=& \underbracket{\int \phi^i_\mathcal{W}(z;k_x) {\rm d}k_x}_{\overline{u^2}_{\mathcal{W}}} + \underbracket{\int \phi_\mathcal{WL}(z;k_x) {\rm d}k_x}_{\overline{u^2}_{\mathcal{WL}}} + \underbracket{\int \phi^c_\mathcal{L}(z;k_x) {\rm d}k_x}_{\overline{u^2}_{\mathcal{L}}}.
\end{eqnarray}
Figure~\ref{fig:TKEdecom}(\emph{b}) presents these three sub-components of the TI at $z^+ \approx 101$, together with $\overline{u^2}(z)$ (open diamonds). Wall-normal profiles of the three sub-components are obtained when this integration is performed for each $z$ (Figure~\ref{fig:TKEdecom}\emph{c}). Note that the contributions are shown in a cumulative format: the bottom profile (squares) represents $\overline{u^2}_{\mathcal{W}}$, the intermediate profile (circles) encompasses $\overline{u^2}_{\mathcal{W}} + \overline{u^2}_{\mathcal{W}\mathcal{L}}$, whereas the final profile (diamonds), $\overline{u^2}_{\mathcal{W}} + \overline{u^2}_{\mathcal{W}\mathcal{L}} + \overline{u^2}_{\mathcal{L}}$, equals $\overline{u^2}$ by construction. Regarding the full $\overline{u^2}(z)$ profile, it is well-known that the near-wall streamwise TI is attenuated due to spatial resolution effects of hot-wires \citep{hutchins:2009a}. Here the spanwise width of the hot-wire sensing length was $l^+ \approx 22$. A corrected profile for the streamwise TI is superposed in Figure~\ref{fig:TKEdecom}(\emph{b}) with filled diamonds, following the method of \citet{smits:2011a}. \citet{samie:2018a} confirmed that this correction scheme is valid for Reynolds numbers up to $Re_\tau \approx 20\,000$. Because the TI above the near-wall region (say $z > z_T$) is unaffected by spatial resolution issues, we proceed our analysis without hot-wire corrections. 

The wall-incoherent component, $\overline{u^2}_{\mathcal{W}}$, exhibits an increase in its energy-magnitude with increasing $z$ throughout the logarithmic region. Section~\ref{sec:empmodel} addresses the wall-normal trend of this TI component in more detail.
\begin{figure} 
\vspace{10pt}
\centering
\includegraphics[width = 0.999\textwidth]{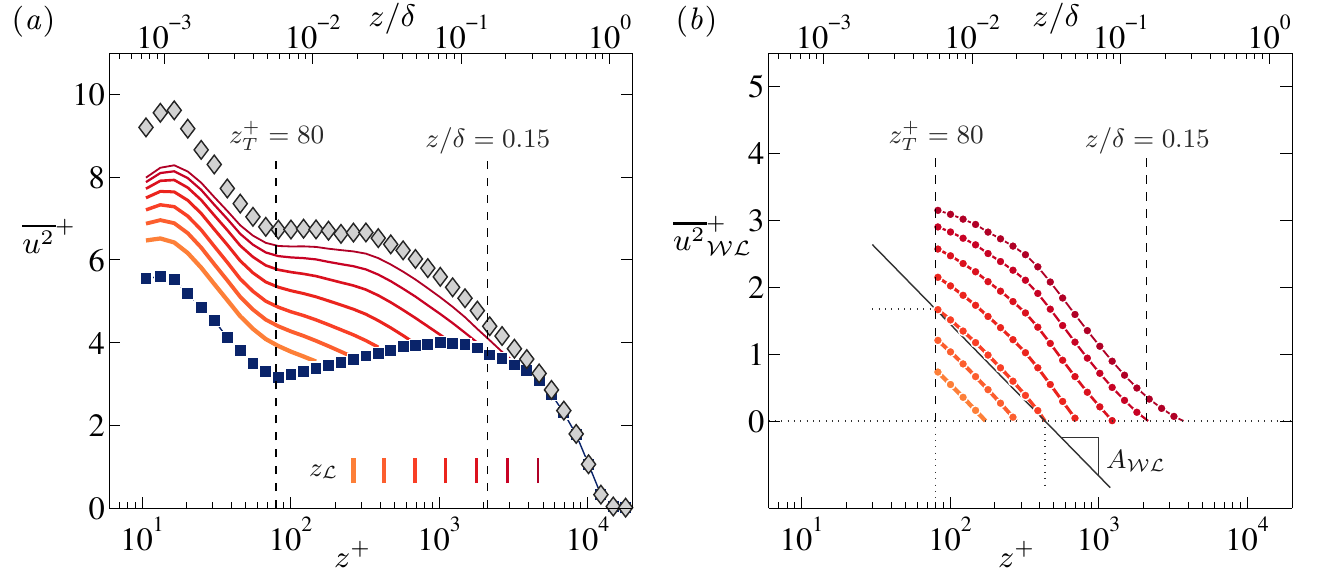}
   \caption{(\emph{a}) Streamwise TI at $Re_\tau \approx 14\,100$. Component $\overline{u^2}_{\mathcal{W}}$ is shown with the square symbols, while the sequence of lines with increasing colour intensity represents $\overline{u^2}_{\mathcal{W}} + \overline{u^2}_{\mathcal{W}\mathcal{L}}$ for increasing $z_{\mathcal{L}}$; locations of $z_\mathcal{L}$ are indicated with the vertical lines. (\emph{b}) Profiles of $\overline{u^2}^+_{\mathcal{W}\mathcal{L}}$.}
   \label{fig:AEprofs}
\end{figure}
Components $\overline{u^2}_{\mathcal{W}\mathcal{L}}$ and $\overline{u^2}_\mathcal{L}$ have to be considered in relation to one another. Figure~\ref{fig:AEprofs}(\emph{a}) illustrate the dependence of the two TI sub-components on $z_\mathcal{L}$, by presenting $\overline{u^2}_\mathcal{W}$ (squares) and $\overline{u^2}_\mathcal{W} + \overline{u^2}_{\mathcal{W}\mathcal{L}}$ (lines) for a range of $z_\mathcal{L}$ (indicated with the vertical lines). Part~1 addressed how their spectral-equivalents, $\phi_{\mathcal{W}\mathcal{L}}$ and $\phi^c_\mathcal{L}$, varied with $z_\mathcal{L}$ and here it is described what implications that has on the streamwise TI. At low $z_\mathcal{L}$, the wall-attached motions smaller than $z = z_\mathcal{L}$ contribute to $\phi_{\mathcal{W}\mathcal{L}}$, but its wall-normal range is limited (per definition, $\phi_{\mathcal{W}\mathcal{L}}$ is non-existent above $z_\mathcal{L}$). With increasing $z_\mathcal{L}$, the range of wall-attached motions increases, but global modes (or imprints of non-self-similar VLSMs/superstructures) that are restricted to $z < z_\mathcal{L}$ also start to contribute significantly to $\phi_{\mathcal{W}\mathcal{L}}$ (due to the inherent difficulty in spectrally decomposing the two, see \S\,5.2 in Part~1). Hence, $\phi_{\mathcal{W}\mathcal{L}}$ does not just contain energy from wall-attached self-similar motions. When $z_\mathcal{L}$ resides in the intermittent region, all global modes are being assigned to $\phi_{\mathcal{W}\mathcal{L}}$ (and thus to $\overline{u^2}_{\mathcal{W}\mathcal{L}}$). This is reflected by the highest $z_\mathcal{L}$ profile in Figure~\ref{fig:AEprofs}(\emph{a}): in the process of increasing $z_\mathcal{L}$, a hump has appeared in the streamwise TI (approaching $\overline{u^2}$ for $z_\mathcal{L} \rightarrow \delta$).

We now focus exclusively on $\overline{u^2}_{\mathcal{W}\mathcal{L}}$ as this sub-component is closely aligned with the scaling following (\ref{eq:u2log}). Figure~\ref{fig:AEprofs}(\emph{b}) shows $\overline{u^2}_{\mathcal{W}\mathcal{L}}$ for $z^+ > z^+_T$ (the near-wall TI is irrelevant in this discussion). Although it was pointed out above that wall-normal profiles of $\overline{u^2}_{\mathcal{W}\mathcal{L}}$ do comprise a signature of wall-attached non-self-similar motions, two trends of its statistics are reflective of wall-attached self-similar motions:\\[-8pt]
\begin{figure} 
\vspace{10pt}
\centering
\includegraphics[width = 0.999\textwidth]{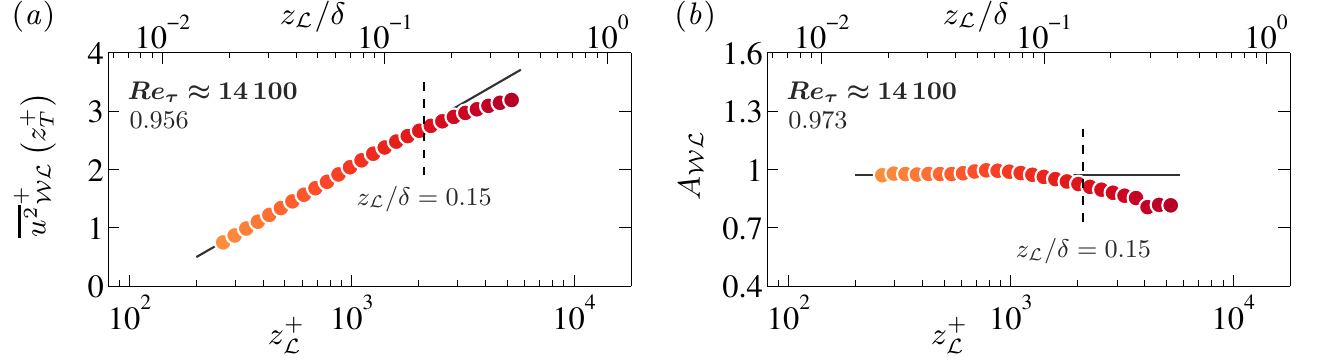}
   \caption{(\emph{a}) $\overline{u^2}^+_{\mathcal{W}\mathcal{L}}$ at $z^+_T = 80$ from Figure~\ref{fig:AEprofs}(\emph{b}); the line, following (\ref{eq:AEslope1}), is fit to the data at $z_\mathcal{L} < 0.15\delta$ (its slope $A'_{\mathcal{W}\mathcal{L}}$ is listed). (\emph{b}) $A_{\mathcal{W}\mathcal{L}}$, as indicated in Figure~\ref{fig:AEprofs}(\emph{b}), superposed on the line corresponding to its mean for data at $z_\mathcal{L} < 0.15\delta$ (value listed).}
   \label{fig:AEslope}
\end{figure}
\begin{enumerate}[labelwidth=0.65cm,labelindent=0pt,leftmargin=0.65cm,label=(\roman*),align=left]
\item \noindent At first, the magnitude of $\overline{u^2}_{\mathcal{W}\mathcal{L}}$ at $z_T$ is displayed in Figure~\ref{fig:AEslope}(\emph{a}), with $z_\mathcal{L}$ forming the abscissa (with a finer $z_\mathcal{L}$-discretization than used in Figure~\ref{fig:AEprofs}). When assuming that the non-self-similar, large-scale motions have a negligible influence on the TI-trend at $z_T$, and that $\overline{u^2}_{\mathcal{W}\mathcal{L}}$ obeys Townsend's AEH, we arrive at
\begin{eqnarray}
 \label{eq:AEslope1}
 \overline{u^2}_{\mathcal{W}\mathcal{L}} \propto A'_{\mathcal{W}\mathcal{L}}\ln\left(\frac{z_\mathcal{L}}{\delta}\right).
\end{eqnarray}
That is, an increase of $z_\mathcal{L}$ mimics an increase in $Re_\tau$ through the inclusion of more wall-attached scales in $\overline{u^2}_{\mathcal{W}\mathcal{L}}$. Data in Figure~\ref{fig:AEslope}(\emph{a}) adheres to (\ref{eq:AEslope1}) for approximately one decade in $z_\mathcal{L}$ and fitting of the data at $z_\mathcal{L} < 0.15\delta$ results in $A'_{\mathcal{W}\mathcal{L}} \approx 0.956$ (note that $z_\mathcal{L} = 0.15\delta$ is often taken as the upper edge of the logarithmic region).
\item \noindent A second measure that is reflective of self-similar wall-attached motions is the decay of $\overline{u^2}_{\mathcal{W}\mathcal{L}}(z)$ following (\ref{eq:u2log}), which is now used to quantify the trend in $\overline{u^2}_{\mathcal{W}\mathcal{L}}$. It is impossible to perform a direct fit of a logarithmic decay to the data of $\overline{u^2}_{\mathcal{W}\mathcal{L}}$ in Figure~\ref{fig:AEprofs}(\emph{b}), because of the aforementioned issues (for large $z_\mathcal{L}$, the profiles are influenced by non-self-similar, global-mode turbulence). Instead, a logarithmic slope $A_{\mathcal{W}\mathcal{L}}$ is defined from the two profile end-points, $\overline{u^2}_{\mathcal{W}\mathcal{L}}(z_T)$ and $\overline{u^2}_{\mathcal{W}\mathcal{L}} = 0$, via
\begin{eqnarray}
 \label{eq:AEslope2}
 A_{\mathcal{W}\mathcal{L}} \equiv \frac{\overline{u^2}_{\mathcal{W}\mathcal{L}}\left(z^+_T\right)}{\ln\left(\left. z^+\right\vert_{\overline{u^2}_{\mathcal{W}\mathcal{L}} = 0}\right) - \ln\left(z^+_T\right)}.
\end{eqnarray}
Figure~\ref{fig:AEprofs}(\emph{b}) displays the logarithmic slope for one profile (discrete point measurements were interpolated to exactly $z^+_T = 80$ and the $z^+$ position at which $\overline{u^2}_{\mathcal{W}\mathcal{L}}$ becomes zero). Data in Figure~\ref{fig:AEslope}(\emph{b}), and its mean value $A_{\mathcal{W}\mathcal{L}} \approx 0.973$, are in close agreement to $A'_{\mathcal{W}\mathcal{L}} \approx 0.956$ from Figure~\ref{fig:AEslope}(\emph{a}). This is expected when $\overline{u^2}_{\mathcal{W}\mathcal{L}}$ obeys an attached-eddy scaling.\\[-8pt]
\end{enumerate}

\subsection{Reynolds number variation}\label{sec:scalingRe}
\begin{figure} 
\vspace{10pt}
\centering
\includegraphics[width = 0.999\textwidth]{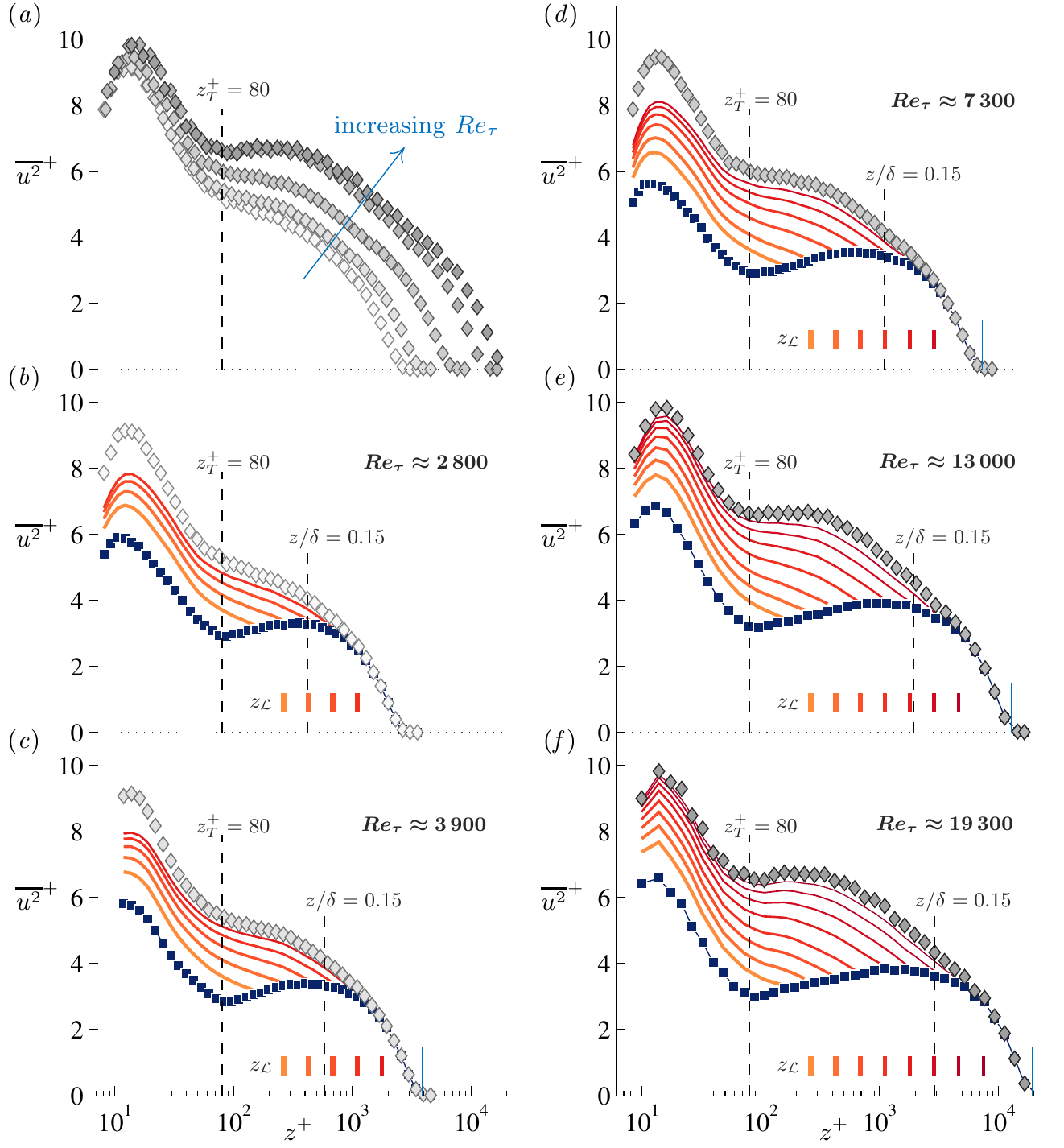}
   \caption{(\emph{a}) Streamwise TI profiles for $Re_\tau \approx 2\,800$, 3\,900 and 7\,300 \citep{hutchins:2009a} and $Re_\tau \approx 13\,000$ and 19\,000 \citep{samie:2018a}. (\emph{b}-\emph{f}) Profiles of (a) decomposed into various TI sub-components: $\overline{u^2}_\mathcal{W}$ is shown with the square symbols, while the sequence of lines with increasing colour intensity represents $\overline{u^2}_\mathcal{W} + \overline{u^2}_{\mathcal{W}\mathcal{L}}$ for increasing $z_\mathcal{L}$ (similar to Figure~\ref{fig:AEprofs}\emph{a}).}
   \label{fig:AEprofsRe}
\end{figure}

We now assess how the identified logarithmic scalings via (\ref{eq:AEslope1}) and (\ref{eq:AEslope2}) depend on the Reynolds number. Single-point hot-wire measurements at a range of Reynolds numbers were employed in \S\,6 of Part~1 to address the Reynolds number variation of the triple-decomposed energy spectrograms. These same single-point hot-wire data are here processed via the procedure described previously (\S\,\ref{sec:scaling}). At first, the $\overline{u^2}(z)$ profiles for these data are shown in Figure~\ref{fig:AEprofsRe}(\emph{a}). For the three lowest Reynolds numbers \citep[$Re_\tau \approx 2\,800$, 3\,900 and 7\,300:][]{hutchins:2009a}, data were corrected for spatial attenuation effects \citep{smits:2011a}, whereas the two other profiles \citep[$Re_\tau \approx 13\,000$ and 19\,300:][]{samie:2018a} comprise fully-resolved measurements. An energy-growth in the outer region presents itself through the emergence of a local maximum in $\overline{u^2}$ \citep{samie:2018a}, whereas at the same time, the near-wall TI grows with $Re_\tau$ \citep{marusic:2017a}. 

\begin{figure} 
\vspace{10pt}
\centering
\includegraphics[width = 0.999\textwidth]{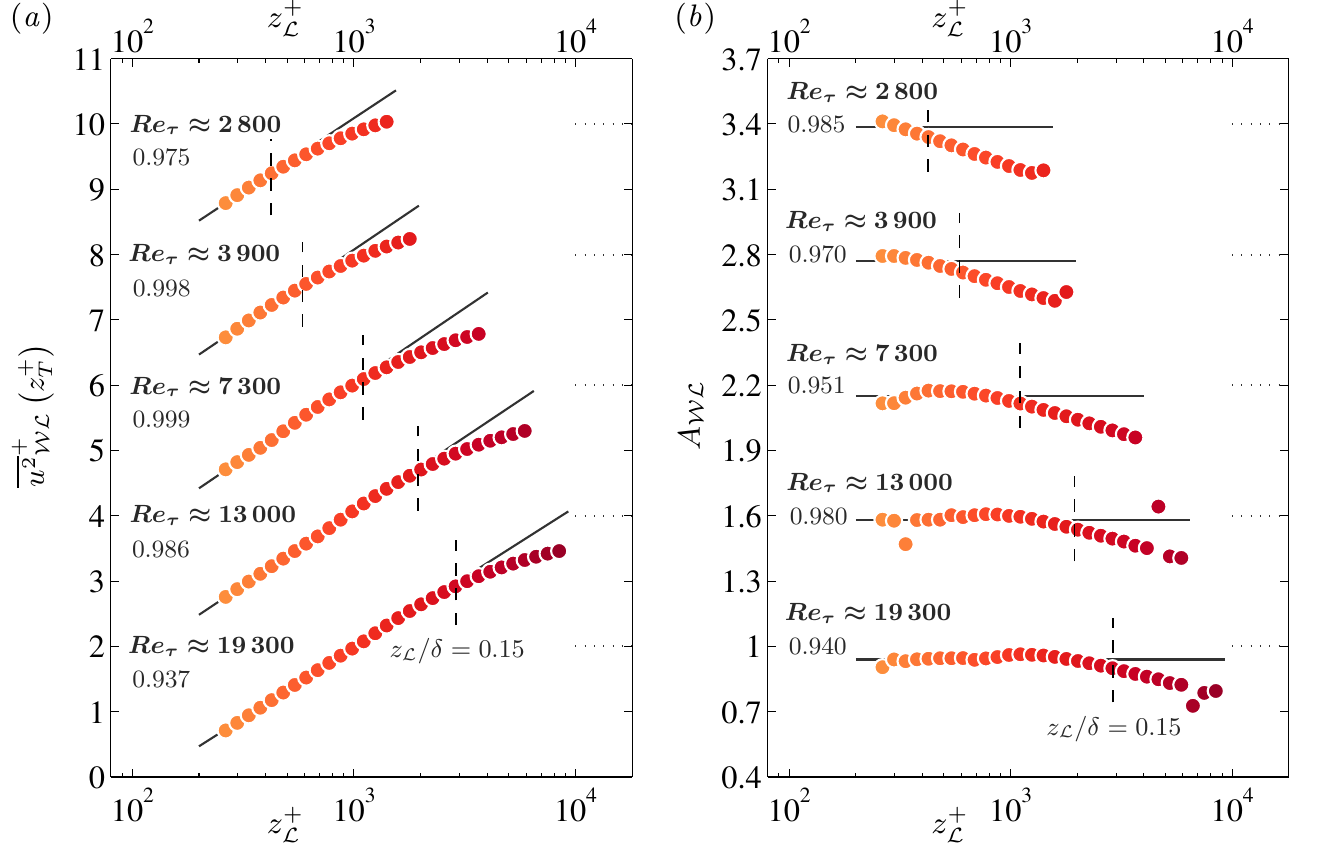}
   \caption{(\emph{a}) $\overline{u^2}^+_{\mathcal{W}\mathcal{L}}$ at $z^+_T = 80$ from Figures~\ref{fig:AEprofsRe}(\emph{b}-\emph{f}); lines, following (\ref{eq:AEslope1}), are fit to the data at $z_\mathcal{L} < 0.15\delta$ (their slopes $A'_{\mathcal{W}\mathcal{L}}$ are listed). (\emph{b}) $A_{\mathcal{W}\mathcal{L}}$, superposed on the line corresponding to its mean value for data at $z_\mathcal{L} < 0.15\delta$ (values listed). Each subsequent $Re_\tau$ case is vertically offset by 2 and 0.6, starting with the second from the bottom, in (\emph{a}) and (\emph{b}), respectively.}
   \label{fig:AEslopeRe}
\end{figure}

Data of each Reynolds-number case are spectrally decomposed to generate a similar output as presented in Figure~\ref{fig:AEprofs}(\emph{a}). For each of the five $Re_\tau$ profiles in Figure~\ref{fig:AEprofsRe}(\emph{a}), the result is shown in Figures~\ref{fig:AEprofsRe}(\emph{b}--\emph{f}), respectively. Additionally, with the aid of (\ref{eq:AEslope1}) and (\ref{eq:AEslope2}), Figures~\ref{fig:AEslope}(\emph{a},\emph{b}) are constructed for each of the five Reynolds numbers, as shown in Figures~\ref{fig:AEslopeRe}(\emph{a},\emph{b}).

Especially at the two largest Reynolds numbers ($Re_\tau \approx 13\,000$ and 19\,300), there is a consistent agreement between $A_{\mathcal{W}\mathcal{L}}$ and $A'_{\mathcal{W}\mathcal{L}}$, which is indicative of the slopes being a reflection of attached-eddy type turbulence. At the two lowest Reynolds numbers ($Re_\tau \approx 2\,800$ and 3\,900), the $A_{\mathcal{W}\mathcal{L}}$ slope extracted from the two profile end-points of $\overline{u^2}_{\mathcal{W}\mathcal{L}}$ exhibits a decreasing trend (top two profiles in Figure~\ref{fig:AEslopeRe}\emph{b}). This is ascribed to the fact that the upward trend of $\overline{u^2}_\mathcal{W}$ (square symbols in Figures~\ref{fig:AEprofsRe}\emph{b},\emph{c}) changes rapidly near the upper edge of the logarithmic region: its magnitude starts to decrease around $z/\delta \approx 0.15$ in order to merge with the TI profiles in the wake region. Because of this decrease, there is a less rapid decay of the $\overline{u^2}_{\mathcal{W}\mathcal{L}}$ profiles near $z/\delta \approx 0.15$. When slope $A_{\mathcal{W}\mathcal{L}}$ is determined from the two profile end-points, it causes a decreased slope. Generally, the limited scale separation in the triple-decomposed spectrograms at low Reynolds numbers exacerbates this issue (see also the spectrograms in Figure~18 of Part~1).
\begin{table} 
  \begin{center}
  \begin{minipage}{\textwidth}  
  \vspace*{-6pt}
  \begin{tabular}{@{}rccccccccc@{}}
  ~ & \multicolumn{6}{l}{Turbulence intensity-based} & ~ & \multicolumn{2}{l}{Spectrum-based} \\\cmidrule{2-7}\cmidrule{9-10}
  $Re_\tau$ & $A'_{\mathcal{W}\mathcal{L}}$ & $A_{\mathcal{W}\mathcal{L}}$ & Part,\S & ~ & $B_1$ & Section & ~ & $k^+_x\phi^+_{uu}\vert_{\rm peak}$ & Section\\[2pt]\cmidrule{2-4}\cmidrule{6-7}\cmidrule{9-10}
  2\,000 & -- & -- & -- & ~ & -- & -- & ~ & 0.195 & 1,\,\S\,1.1\\
  2\,800 & $0.975\pm 0.056$ & $0.985\pm 0.040$ & 2,\,\S\,\ref{sec:scalingRe} & ~ & $-2.224\pm 0.122$ & 2,\,\S\,\ref{sec:A1summ} & ~ & 0.344 & 1,\,\S\,6.2\\
  3\,900 & $0.998\pm 0.041$ & $0.970\pm 0.022$ & 2,\,\S\,\ref{sec:scalingRe} & ~ & $-2.340\pm 0.095$ & 2,\,\S\,\ref{sec:A1summ} & ~ & 0.466 & 1,\,\S\,6.2\\
  7\,300 & $0.999\pm 0.020$ & $0.951\pm 0.014$ & 2,\,\S\,\ref{sec:scalingRe} & ~ & $-2.388\pm 0.053$ & 2,\,\S\,\ref{sec:A1summ} & ~ & 0.685 & 1,\,\S\,6.2\\
  13\,000 & $0.986\pm 0.010$ & $0.980\pm 0.017$ & 2,\,\S\,\ref{sec:scalingRe} & ~ & $-2.290\pm 0.030$ & 2,\,\S\,\ref{sec:A1summ} & ~ & 0.851 & 1,\,\S\,6.2\\
  14\,100 & $0.956\pm 0.013$ & $0.973\pm 0.010$ & 2,\,\S\,\ref{sec:scaling} & ~ & $-2.189\pm 0.038$ & 2,\,\S\,\ref{sec:A1summ} & ~ & 0.900 & 1,\,\S\,6.2\\
  19\,300 & $0.937\pm 0.011$ & $0.940\pm 0.009$ & 2,\,\S\,\ref{sec:scalingRe} & ~ & $-2.168\pm 0.036$ & 2,\,\S\,\ref{sec:A1summ} & ~ & 0.938 & 1,\,\S\,6.2\\[2pt]
  \end{tabular}
  \caption{Values of $A'_{\mathcal{W}\mathcal{L}}$, $A_{\mathcal{W}\mathcal{L}}$ and $B_1$, which were inferred from the $\overline{u^2}_{\mathcal{W}\mathcal{L}}$ profiles. Uncertainty estimates for $A_1 = A'_{\mathcal{W}\mathcal{L}}$ and $B_1$ are based on 95\,\% confidence bounds from the fitting of (\ref{eq:u2logcor}), while the uncertainty estimates for $A_{\mathcal{W}\mathcal{L}}$ are based on the 95\,\% confidence interval of the data points residing at $z_\mathcal{L} < 0.15\delta$ in Figures~\ref{fig:AEslope}(\emph{b}) and~\ref{fig:AEslopeRe}(\emph{b}).}
  \label{tab:summ}
  \end{minipage}
  \end{center}
\end{table}

\subsection{Reconciling $A_1$ from trends in the turbulence intensity and spectra}\label{sec:A1summ}
\begin{figure} 
\vspace{10pt}
\centering
\includegraphics[width = 0.999\textwidth]{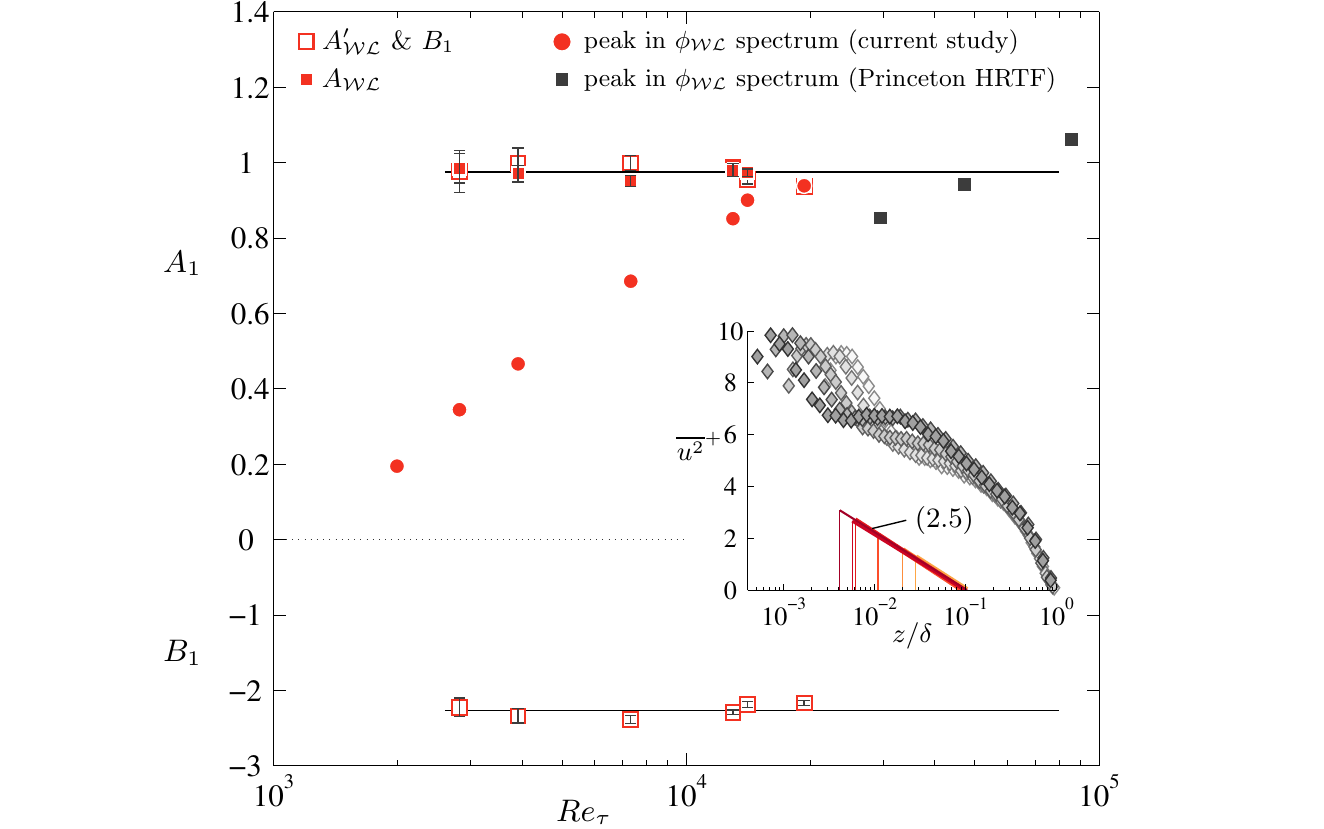}
   \caption{Values of $A'_{\mathcal{W}\mathcal{L}}$ and $A_{\mathcal{W}\mathcal{L}}$, alongside the peak values of $k^+_x\phi^+_{\mathcal{W}\mathcal{L}}$ (duplicated from Figure~20, Part~1). Values of $B_1$ are plotted at the bottom. TI profiles of Figure~\ref{fig:AEprofsRe}(\emph{a}) are shown in the inset, together with (\ref{eq:u2logcor}) for all six cases of $A_1 = A'_{\mathcal{W}\mathcal{L}}$ and $B_1$.}
   \label{fig:A1summ}
\end{figure}

Having re-assessed the wall-normal decay of the TI sub-component associated with Townsend's attached-eddies (\S\S,\ref{sec:scaling}-\ref{sec:scalingRe}), we can now proceed with reconciling the \emph{status quo}. Recall that (\ref{eq:u2log}) is restricted to the streamwise TI that is generated by inviscid, geometrically self-similar and wall-attached eddies only. Because both $A'_{\mathcal{W}\mathcal{L}}$ and $A_{\mathcal{W}\mathcal{L}}$ were inferred by considering the sub-component of the TI that complies with Townsend's assumptions only, those slopes are interpreted as $A_1$. Figure~\ref{fig:A1summ} displays $A'_{\mathcal{W}\mathcal{L}}$, for all Reynolds numbers, with the open square symbols. Uncertainty estimates are shown with the error bars and are based on 95\,\% confidence bounds from the fitting procedure of (\ref{eq:AEslope1}). Alongside, with the solid square symbols, values of $A_{\mathcal{W}\mathcal{L}}$ are shown with the uncertainty estimates based on the 95\,\% confidence interval of the data points residing at $z_\mathcal{L} < 0.15\delta$ in Figures~\ref{fig:AEslope}(\emph{b}) and~\ref{fig:AEslopeRe}(\emph{b}). Numerical values are summarized in Table~1. To complete quantification of (\ref{eq:u2log}) by considering $\overline{u^2}_{\mathcal{W}\mathcal{L}}$ energy only, offset $B_1$ can be determined. For this we have to introduce a new quantity $\overline{u^2}_{\rm AE}$, being the TI decay with a pure logarithmic decay:
\begin{eqnarray}
 \label{eq:u2logcor}
 \overline{u^2}^+_{\rm AE}\left(z/\delta\right) = B_1\left(z_\mathcal{L} = 0.15\delta\right) - A_1\ln\left(\frac{z}{\delta}\right)\text{,~~~for~} z^+ \geqslant z^+_T.
\end{eqnarray}
Although offset $B_1$ depends on $z_\mathcal{L}$ (see Figure~\ref{fig:AEprofs}\emph{b}), we only have to consider the scenario for one specific $z_\mathcal{L}$ to infer its Reynolds-number trend (here we take $z_\mathcal{L} = 0.15\delta$). Values for $B_1(z_\mathcal{L} = 0.15\delta)$ are shown on the bottom of Figure~\ref{fig:A1summ}. Mean values for both $A_1$ and $B_1$ are found from the mean values of $A_1 = A'_{\mathcal{W}\mathcal{L}}$ and $B_1$ in Table~\ref{tab:summ}, resulting in
\begin{eqnarray}
 \label{eq:A1B1avg}
 A_1 = 0.975\text{,}~~~~B_1\left(z_\mathcal{L} = 0.15\delta\right) = -2.267.
\end{eqnarray}
To indicate the effect of the variation in $A_1$ and $B_1$ with $Re_\tau$, six lines according to formulation (\ref{eq:u2logcor}), with the six values of $A_1 = A'_{\mathcal{W}\mathcal{L}}$ and $B_1$ (down to $z^+_T = 80$) are shown in the inset in Figure~\ref{fig:A1summ}, together with the TI profiles of Figure~\ref{fig:AEprofsRe}(\emph{a}). The scatter in $A_1$ and $B_1$ (as well as the uncertainty estimates from the fitting procedure, listed in Table~1), result in indistinguishable logarithmic trends in relation to typical experimental uncertainty in the TI profiles \citep[\emph{e.g.}][]{winkel:2012a,vincenti:2013a,marusic:2017a,orlu:2017a,samie:2018a}. Both $A_1$ and $B_1$ are thus considered to be Reynolds number invariant for $2\,800 \apprle Re_\tau \apprle 19\,300$.

A last set of data points in Figure~\ref{fig:A1summ} comprises the peak values of $k^+_x\phi^+_{\mathcal{W}\mathcal{L}}$ at $z^+ = 100$, duplicated from Figure~20 in Part~1. These data consider $A_1$ in the context of the energy spectra. Because the scale separation in spectral space is still relatively limited at these Reynolds numbers, the peak value in the associated spectra ($k^+_x\phi^+_{\mathcal{W}\mathcal{L}}$) keeps maturing with $Re_\tau$ (detailed in \S\,6.2, Part~1). At $Re_\tau \approx 19\,300$ there is a consistency between the value for $A_1$ found from the TI trend and the peak/plateau in the associated spectrum. However, it is acknowledged that no complete similarity has been observed in the associated $\phi_{\mathcal{W}\mathcal{L}}$ spectra (\emph{e.g.} Figure~19, Part~1). Furthermore, the three Princeton HRTF data points of the peak in the $\phi_{\mathcal{W}\mathcal{L}}$ spectrum are not conclusive on whether the spectral peak plateaus. The difficult conditions in the Princeton HRTF and the challenging aspects associated with the acquisition of repeatable statistics with miniature hot-wire probes \citep[\emph{e.g.}][]{samie:2018a} require more future research to settle this issue. Only fully-resolved, higher Reynolds number data at $z^+ = \mathcal{O}(100)$ can provide definitive answers on whether a distinguished $k_x^{-1}$ region develops (in which case the peak has to plateau to a Reynolds number-invariant $A_1$ consistent with the TI trends). In the interim, Figure~\ref{fig:A1summ} does not exclude that possibility: a rough extrapolation of the $\phi_{\mathcal{W}\mathcal{L}}$ peak values approaches $A_1 \approx 1$, consistent with (\ref{eq:A1B1avg}). Finally, to the authors' knowledge, our current work indicates for the first time that a Reynolds number-invariant $A_1$ could be consistent with a potential $k_x^{-1}$ at ultra high $Re_\tau$. Previously, $A_1 = 1.03$ \citep{perry:1990a,marusic:2003a} found from $\overline{u^2}(z)$ profiles was in close agreement with the spectral-based value of $A_1 \approx 0.95$ by \citet{nickels:2005a}, but this was strictly coincidental. The former was obtained at significantly lower Reynolds numbers (highest $Re_\tau \approx 4\,400$) than the latter ($Re_\tau \approx 14\,000$).


\section{$A_1$ in relation to the turbulence intensity in the near-wall region}\label{sec:nearwall}
\begin{figure} 
\vspace{10pt}
\centering
\includegraphics[width = 0.999\textwidth]{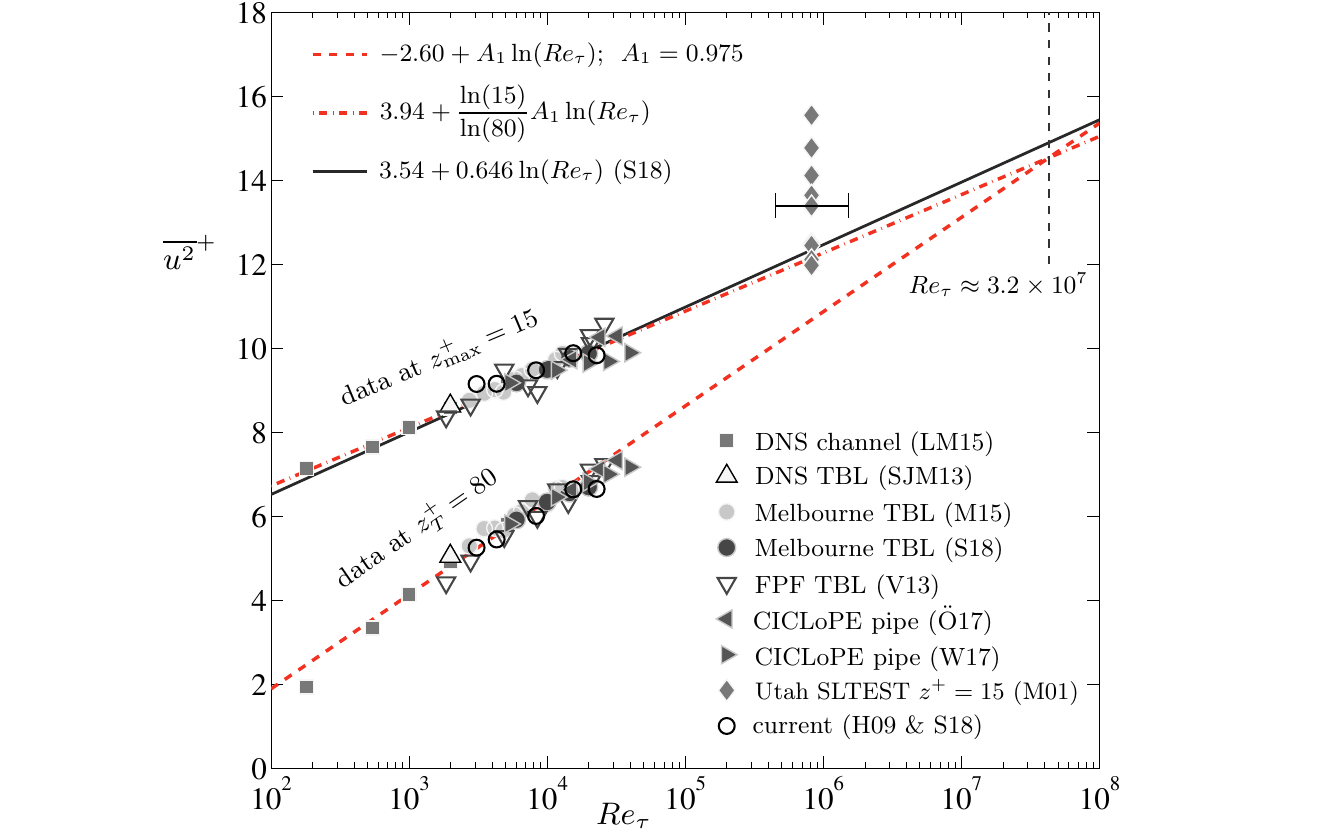}
   \caption{Turbulence intensity, $\overline{u^2}$, at the nominal wall-normal location of the inner-peak, $z^+_{\rm max} = 15$, and at $z^+_T = 80$. For the $z^+_T$ data, a dashed line shows a Reynolds-number growth according to the attached-eddy scaling with $A_1 = 0.975$ (\ref{eq:A1B1avg}). For the $z^+_{\rm max}$ data, the dash-dotted line has a slope of $\ln(15)/\ln(80)A_1 = 0.603$, while the solid line represents (\ref{eq:samie}). Data are from LM15: \citet{lee:2015a}, SJM13: \citet{sillero:2013a}, M15: \citet{marusic:2015a}, S18: \citet{samie:2018a}, V13: \citet{vincenti:2013a}, \"{O}17: \citet{orlu:2017a}, W17: \citet{willert:2017a}, M01: \citet{metzger:2001a} and H09: \citet{hutchins:2009a}.}
   \label{fig:TKEwalldata}
\end{figure}
Consistent scaling laws have recently emerged for the inner-peak of the streamwise TI. \citet{samie:2018a} considered the maximum in the TI profiles, denoted as $\overline{u^2}_{\rm max}$, from DNS and fully-resolved measurement data, to conclude that
\begin{eqnarray}
 \label{eq:samie}
 \overline{u^2}^+_{\rm max} = C + A_{\rm max}\ln\left(Re_\tau\right),
\end{eqnarray}
with $C = 3.54$ and $A_{\rm max} = 0.646$. Nominally, the maxima reside at $z^+_{\rm max} = 15$. \citet{lee:2015a} observed (\ref{eq:samie}) through DNS channel flow up to $Re_\tau \approx 5\,200$ and an increase of $\overline{u^2}^+_{\rm max}$ with $Re_\tau$ is consistent with earlier studies \citep{degraaff:2000a,hutchins:2009a,klewicki:2010a}. Peak values of the streamwise TI at $z_{\rm max}$, as a function of $Re_\tau$, are shown in Figure~\ref{fig:TKEwalldata}. DNS data include the channel flow of \citet{lee:2015a} and TBL flow of \citet{sillero:2013a}. Experimental data of TBLs are from studies performed in Melbourne's boundary layer facility \citep{marusic:2015a,samie:2018a}, UNH's Flow Physics Facility \citep{vincenti:2013a} and at Utah SLTEST \citep{metzger:2001a}. Current data (employed in Figure~\ref{fig:AEslopeRe}). All these data \citep[aside from][]{samie:2018a} were corrected for spatial resolution effects via \citet{smits:2011a}. ASL data of \citet{metzger:2001a}, with a relatively small hotwire length of $l^+ \apprle 10$, are uncorrected \citep{hutchins:2009a}. High Reynolds number experimental pipe flow data are also included from the CICLoPE facility \citep{orlu:2017a,willert:2017a}, reaching up to $Re_\tau \approx 40\,000$. Hotwire data of \citet{orlu:2017a} were again corrected for spatial resolution effects, whereas the PIV data of \citet{willert:2017a} were nearly fully-resolved. Given the measurement uncertainty, (\ref{eq:samie}) appears to represent the trend well for all the data (solid line).

Figure~\ref{fig:TKEwalldata} also presents $\overline{u^2}$ at $z^+_T = 80$, except for the unavailable Utah SLTEST data at this location. When the data at $z^+_T = 80$ adhere to an attached-eddy scaling, the Reynolds-number growth of the streamwise TI can be described by $A_1$, since (\ref{eq:u2log}) or (\ref{eq:u2logcor}) can be reformulated as
\begin{eqnarray}
 \label{eq:Relogstart}
 \overline{u^2}^+\left(z^+_T\right) = D + A_1\ln\left(Re_\tau\right).
\end{eqnarray}
When fitting (\ref{eq:Relogstart}) to the data in Figure~\ref{fig:TKEwalldata} with $A_1 = 0.975$ (\ref{eq:A1B1avg}), the offset-constant $D$ is determined as $D = -2.60$. Figure~\ref{fig:TKEwalldata} shows that (\ref{eq:Relogstart}) represents the data well, meaning that the Reynolds-number behaviour of the streamwise TI, at a lower bound of the logarithmic region fixed in viscous scaling, \emph{e.g.}, $80\nu/U_\tau$, is predicted well through an attached-eddy scaling alone. This thus implies that energy footprints from large-scale, global-mode VLSMs and small-scale wall-incoherent turbulence (reflected in $\overline{u^2}_{\mathcal{L}}$ and $\overline{u^2}_\mathcal{W}$, respectively) do not, or negligibly, contribute to the Reynolds-number trend over the range of $Re_\tau$ investigated here. That is, an energetic footprint is still present (clearly observed in component $\phi^c_\mathcal{L}$ in Figure~2\emph{c}, for instance), but its Reynolds number trend seems weak as an attached-eddy scaling of self-similar turbulence alone can explain the growth of the streamwise TI at $z_T$ and $z^+ = 15$. Interactions between the outer- and inner-region turbulence, however, are not insignificant. They are most pronounced in the near-wall region due to the co-existence of near-wall turbulence and energetic footprints of larger-scale, wall-attached outer motions \citep[\emph{e.g.}][]{marusic:2010a,cho:2018a}.

The question now remains how (\ref{eq:samie}) and (\ref{eq:Relogstart}) are compatible (or how $A_{\rm max} = 0.646$ is consistent with $A_1 = 0.975$). \citet{marusic:2003a} proposed that the near-wall viscous region is influenced by the Reynolds number dependent, outer-layer streamwise TI. The validity of this proposition was strengthened by the superposition framework detailed in the literature \citep{hutchins:2007ab,marusic:2010a,mathis:2011a,baars:2016ab} and studies focusing on a near-wall component that is free of motions not scaling in inner units \citep[\emph{e.g.}][]{hu:2018a}. Note however that a complex scale interaction and spectral energy transfer is present \citep{degiovanetti:2017a,cho:2018a}, in combination with an outer motion wall-shear-stress footprint \citep{abe:2004a,giovanetti:2016a}. In summary, we move forward with the near-wall TI being composed of two contributions:\\[-8pt]
\begin{enumerate}[labelwidth=0.65cm,labelindent=0pt,leftmargin=0.65cm,label=(\roman*),align=left]
\item \noindent A universal function that is Reynolds-number invariant when scaled in inner units, denoted as $\overline{u^2}^+_{\rm NW}(z^+)$. It mainly encompasses the inner-peak in the spectrogram induced by the near-wall cycle (NW cycle), but also comprises a contribution---that seems to be Reynolds-number invariant---from the largest, outer-region motions (see our discussion above).
\item \noindent An additive component that accounts for the Reynolds-number dependent superposition of the outer region TI onto the near-wall viscous region. It is hypothesized that this Reynolds-number dependence is solely the result of the attached-eddy turbulence at $z^+_T = 80$. In simplest form, it can be hypothesized that the near-wall footprint drops off linearly in $\ln(z^+)$, to zero at $z^+ = 1$, so that
\begin{eqnarray}
 \label{eq:super}
 \overline{u^2}^+_{\rm AE}\left(z^+,Re_\tau\right) = \frac{\ln\left(z^+\right)}{\ln\left(z^+_T\right)}\overline{u^2}^+_{\rm AE}\left(z^+_T\right)\text{,~~~for~} 1 \leqslant z^+ \leqslant z^+_T,
\end{eqnarray}
where $\overline{u^2}^+_{\rm AE}\left(z^+_T\right)$ is found from (\ref{eq:u2logcor}), which is reformulated as
\begin{eqnarray}
 \label{eq:uzT}
 \overline{u^2}^+_{\rm AE}\left(z^+_T\right) = \left[B_1(z_\mathcal{L} = 0.15\delta) - A_1\ln\left(z^+_T\right)\right] + A_1\ln\left(Re_\tau\right).
\end{eqnarray}\\[-8pt]\end{enumerate}
\begin{figure} 
\vspace{10pt}
\centering
\includegraphics[width = 0.999\textwidth]{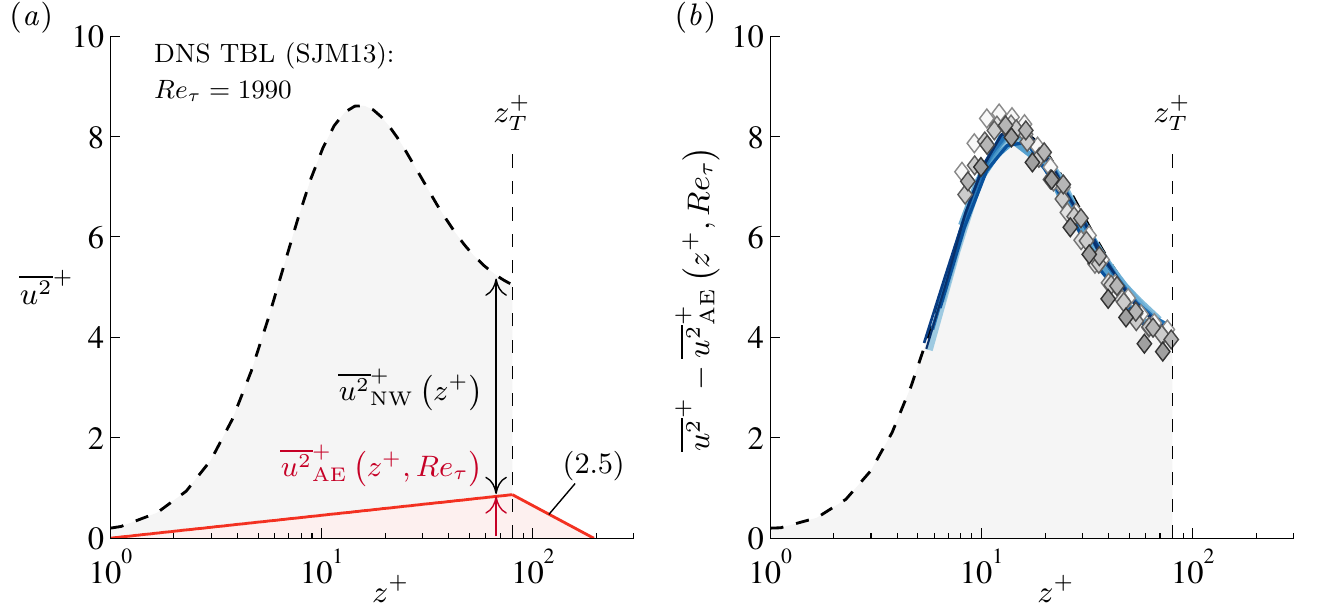}
   \caption{(\emph{a}) Streamwise TI profile of at $Re_\tau = 1990$ \citep[DNS,][]{sillero:2013a}, together with (\ref{eq:super}) and (\ref{eq:u2logcor}). (\emph{b}) Streamwise TI profiles in the near-wall region, with the superposition component of the attached-eddies removed. Dashed line: DNS profile of (\emph{a}), markers: data from Figure~\ref{fig:AEslopeRe}(\emph{a}), blue-coloured lines: 10 profiles from Figure~7 in \citet{marusic:2017a}, spanning $2\,800 < Re_\tau < 13\,400$.}
   \label{fig:TKEnearwall}
\end{figure}
As $\overline{u^2}^+_{\rm NW}(z^+)$ is Reynolds number-invariant, $\overline{u^2}^+(z^+_{\rm max})$ grows with $Re_\tau$ via (\ref{eq:samie}) with $A_{\rm max} = \ln(z^+_{\rm max})/\ln(z^+_T)A_1 = 0.603$. Refitting of (\ref{eq:samie}) yields $C = 3.94$; Figure~\ref{fig:TKEwalldata} indicates that these constants represent the scattered data equally well as with $A_{\rm max} = 0.646$ and $C = 3.54$ \citep[adopted earlier from][]{samie:2018a}. In order to extend the scaling validation to the entire near-wall region (not just $z_{\rm max}$), reference DNS data of a ZPG TBL are utilized \citep{sillero:2013a}. Figure~\ref{fig:TKEnearwall}(\emph{a}) displays $\overline{u^2}$ from the DNS at $Re_\tau = 1990$. Following (\ref{eq:super}), part of this near-wall TI is envisioned as the attached-eddy component, labelled as $\overline{u^2}_{\rm AE}$. The remaining TI forms $\overline{u^2}_{\rm NW}$. For any $Re_\tau$, data must now collapse when the near-wall attached-eddy contribution is subtracted from the near-wall TI profile. Figure~\ref{fig:TKEnearwall}(\emph{b}) visualizes this assessment: the dashed line corresponds to $\overline{u^2}_{\rm NW}$ from the DNS in Figure~\ref{fig:TKEnearwall}(\emph{a}), the symbols correspond to the five Reynolds number profiles of Figure~\ref{fig:AEprofsRe}(\emph{a}) and the 10 blue-coloured profiles span $2\,800 < Re_\tau < 13\,400$ \citep[taken from][where data were also corrected for the hotwire's spatial attenuation effects]{marusic:2017a}. The excellent collapse of all data agrees with the two-part model, $\overline{u^2}(z^+,Re_\tau) = \overline{u^2}_{\rm NW}(z^+) + \overline{u^2}_{\rm AE}(z^+,Re_\tau)$. In conclusion, $A_1 = 0.975$ is consistent with the Reynolds-number increase of the near-wall TI.

\section{Empirical trends in the decomposed turbulence intensity}\label{sec:empmodel}
\begin{figure} 
\vspace{10pt}
\centering
\includegraphics[width = 0.999\textwidth]{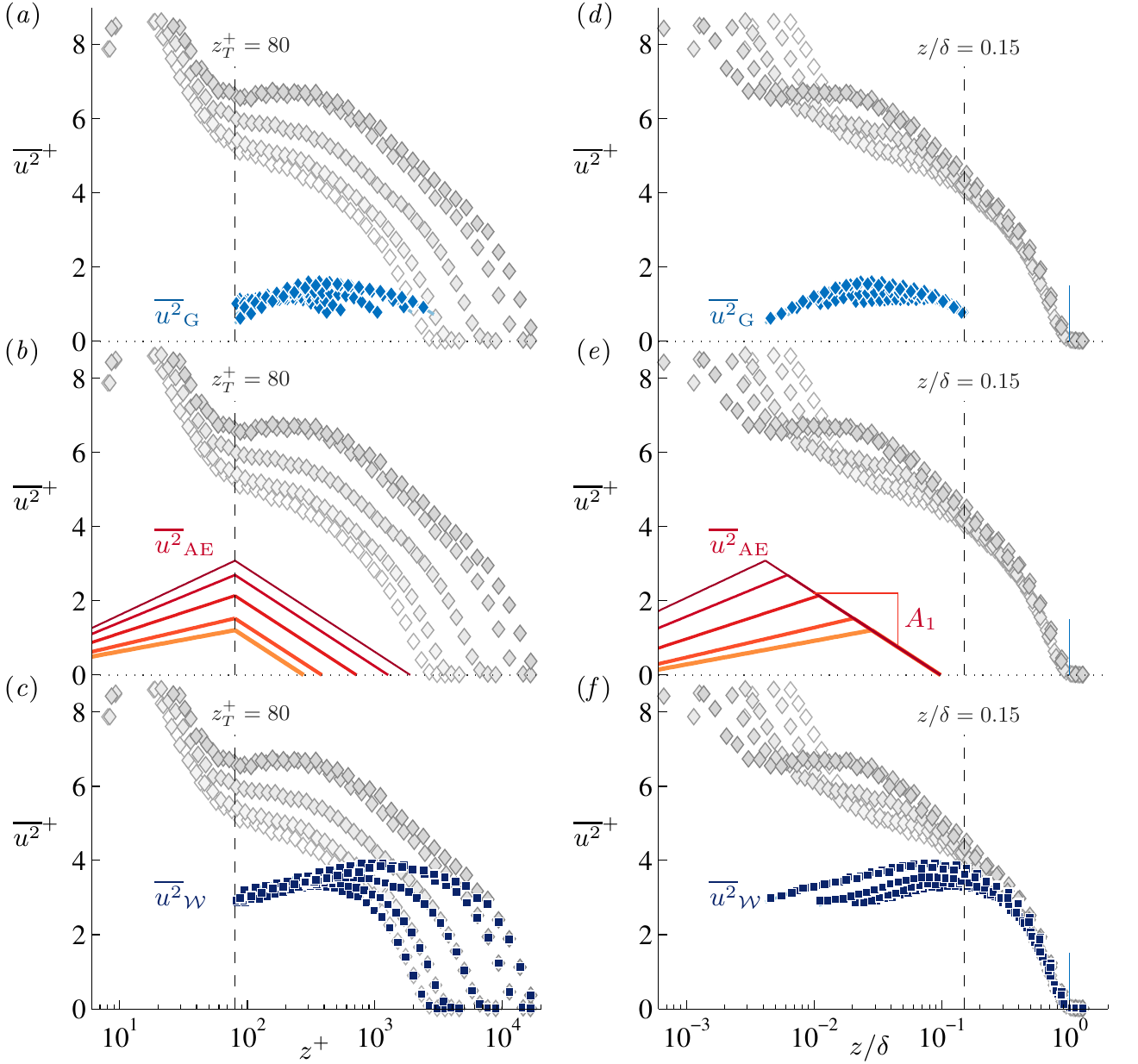}
   \caption{Streamwise TI profiles at a range of Reynolds numbers. Measured TI profiles are shown in each sub-figure with light grey (duplicate from Figure~\ref{fig:AEprofsRe}\emph{a}). (\emph{a}-\emph{f}) In each of the three rows, TI profiles are superposed of one of the three sub-components. Inner-scaling and outer-scaling are used in the left and right columns, respectively.}
   \label{fig:TKEtrdecom}
\end{figure}

Evidence for a portion of the streamwise TI in the \emph{logarithmic region} adhering to (\ref{eq:u2log}) was provided in \S\,\ref{sec:scalingO}. Subsequently, \S\,\ref{sec:nearwall} highlighted its consistency with the \emph{near-wall region} scaling trends. In this section we proceed with an examination of the empirical trends in the decomposed TI, based on the explicit assumption that the attached-eddy turbulence obeys a perfect logarithmic decay with $z$. Following the data-driven spectral decomposition, the streamwise TI was earlier analysed in terms of its three additive components, following (\ref{eq:u2decom2}):
\begin{eqnarray}
 \label{eq:decomold}
 \overline{u^2}^+ = \overline{u^2}^+_{\mathcal{W}} + \overline{u^2}^+_{\mathcal{W}\mathcal{L}} + \overline{u^2}^+_{\mathcal{L}}.
\end{eqnarray}
Wall-coherent components $\overline{u^2}_{\mathcal{W}\mathcal{L}}$ and $\overline{u^2}_\mathcal{L}$, computed by the data-driven approach, are not separable in the sense that one of these consists of attached-eddy turbulence only (the inherent difficulty of decomposing energy wall-coherent self-similar attached-eddies from that of wall-coherent, large-scale non-self-similar motions was discussed in \S\,5.2 of Part~1). However, it was shown that $\overline{u^2}_{\mathcal{W}\mathcal{L}}$ closely represents the energy content associated with attached-eddies, by framing both (\ref{eq:u2logcor}) and (\ref{eq:A1B1avg}). We now proceed with the explicit assumption that (\ref{eq:u2logcor}) exists and investigate the implications of this on the scaling of the sub-components, still totaling the non-decomposed TI, $\overline{u^2}$. First we replace $\overline{u^2}_{\mathcal{W}\mathcal{L}}$ with $\overline{u^2}_{\rm AE}$ (\emph{e.g.}, $\overline{u^2}_{\rm AE}$ obeys pure attached-eddy scaling). Consequently, $\overline{u^2}_\mathcal{L}$ needs to be replaced by a component encompassing all remaining energy, denoted as $\overline{u^2}_{\rm G}$, where subscript G stands for global. Wall-incoherent component $\overline{u^2}_\mathcal{W}$ remains unchanged, resulting in
\begin{eqnarray}
 \label{eq:decomnew}
 \overline{u^2}^+ = \overline{u^2}^+_{\mathcal{W}} + \overline{u^2}^+_{\rm AE} + \overline{u^2}^+_{\rm G}.
\end{eqnarray}
For the data considered in Figure~\ref{fig:AEslopeRe}, with $z_\mathcal{L} = 0.15\delta$, the three additive contributions of (\ref{eq:decomnew}) are displayed in Figure~\ref{fig:TKEtrdecom}(\emph{a}-\emph{c}) and Figures~\ref{fig:TKEtrdecom}(\emph{b}-\emph{f}) for inner- and outer-scalings, respectively. The scaling of each component, in the logarithmic region, is now discussed.

\begin{figure} 
\vspace{10pt}
\centering
\includegraphics[width = 0.999\textwidth]{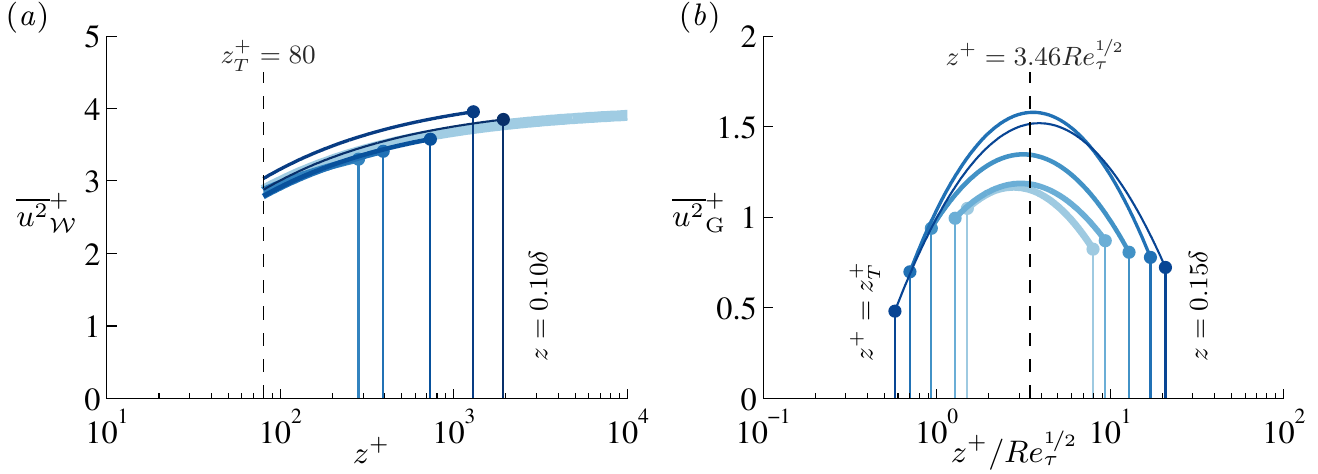}
   \caption{(\emph{a}) Function fits via (\ref{eq:K}) to $\overline{u^2}_{\mathcal{W}}$ in Figure~\ref{fig:TKEtrdecom}(\emph{c}) over the range $z_T \leqslant z \leqslant 0.10\delta$. The thick light blue line in the background presents (\ref{eq:K}) with the average values of $K_1 = 4.01$ and $K_2 = 10.13$. (\emph{b}) Curve fits via (\ref{eq:G}) to $\overline{u^2}_{\rm G}$ in Figure~\ref{fig:TKEtrdecom}(\emph{a}) over the range $z_T \leqslant z \leqslant 0.15\delta$.}
   \label{fig:KGfit}
\end{figure}

The simplest approach for obtaining a scaling formulation for $\overline{u^2}_\mathcal{W}(z) = \int \phi^i_\mathcal{W}(z;k_x) {\rm d}k_x$ is to use Kolmogorov-type modelling as used in previous works \citep[see][]{spalart:1988a,marusic:1997a}. Spectral scaling of $\phi^i_\mathcal{W}$ comprises a $z$-scaling at the low wavenumber-end, while the higher wavenumber-end adheres to a $k^{-5/3}$ scaling up to a wavenumber fixed in Kolmogorov scale $\eta$, see Figures 15(\emph{f}) and~21 in Part~1. When integrating a $k_x^{-5/3}$ model spectrum from $k_xz = c_1$ to $k_xz = c_2z^{3/4}$, where $c_1$ and $c_2$ are constants, a scaling trend can be inferred. The latter boundary equals $k_x\eta = M$, with $M$ being a constant and $\eta \equiv (\nu^3/\epsilon)^{\nicefrac{1}{4}}$. From the $\epsilon \sim 1/z$ production-dissipation balance, we find that $k_x\eta = M \rightarrow k_xz = c_2z^{3/4}$. Accordingly,
\begin{eqnarray}
 \label{eq:K}
 \overline{u^2}^+_{\mathcal{W}}\left(z\right) = \int_{k_xz = c_1}^{k_xz = c_2z^{3/4}} \frac{K_0}{\left(k_xz\right)^{5/3}} {\rm d}\left(k_xz\right) = K_1 - \frac{K_2}{\sqrt{z}},
\end{eqnarray}
where $K_0$, $K_1$ and $K_2$ are constants. When $Re_\tau \rightarrow \infty$, $\overline{u^2}^+_{\mathcal{W}}$ will tend towards $K_1$ for large $z^+$. At our practical values of $Re_\tau$ (Figure~\ref{fig:TKEtrdecom}\emph{c}), $\overline{u^2}^+_{\mathcal{W}}$ is seen to increase up to $z \approx 0.10\delta$, after which a wake deviation occurs \citep{marusic:1997a}. Fitting of (\ref{eq:K}) to the data in Figure~\ref{fig:TKEtrdecom}(\emph{c}), for $z_T \leqslant z \leqslant 0.10\delta$, results largely in a Reynolds number-invariant contribution as shown by the profiles in Figure~\ref{fig:KGfit}(\emph{a}); it was visually verified that (\ref{eq:K}) adequately described each experimental profile in Figure~\ref{fig:TKEtrdecom}(\emph{c}). Experimental uncertainties in $U_\tau$ can be a cause for the slight variations observed in Figure~\ref{fig:KGfit}(\emph{a}). Average values for the constants, from the five profiles, were found to be $K_1 = 4.01$ and $K_2 = 10.13$ (thick light blue line in Figure~\ref{fig:KGfit}\emph{a}). By inspection, the data in Figure~\ref{fig:TKEtrdecom}(\emph{c}) is not described by (\ref{eq:K}) above $z \approx 0.10\delta$. This deviation is expected since the $\overline{u^2}^+_{\mathcal{W}}$ is strictly formed from the stochastic, wall-incoherent energy. At sufficiently large $z$, the wall-scaling of filter $f_\mathcal{W}$ breaks down and $\overline{u^2}^+_{\mathcal{W}}$ begins to include all turbulent scales (not just the scales from an inertial sub-range and dissipative-end of the cascade). At the same time, the more limited scale separation in the wake, as well as the effects of intermittency on spectra \citep{kwon:2016a}, make it impossible to physically interpret the results in the wake region.

For the attached-eddy energy $\overline{u^2}_{\rm AE}$, (\ref{eq:u2logcor}) was adopted. The near-wall decay trend following (\ref{eq:super}) is also drawn in Figures~\ref{fig:TKEtrdecom}(\emph{b},\emph{e}). The vertical offset of the AE component, being $B_1$ in (\ref{eq:u2logcor}), is dependent on the chosen $z_\mathcal{L}$. Further research should provide insight into what offset describes the stochastic statistics of attached structures (at what outer-scaled location $\overline{u^2}_{\rm AE}$ should become zero). Conceivably, studies extracting instantaneous attached-eddy structures from full velocity fields can be instrumental to this \citep{delalamo:2006a,lozano:2012a,hwang:2018a,solak:2018a}.

Finally, $\overline{u^2}_{\rm G}$ appears as a broad hump throughout the logarithmic region (Figures~\ref{fig:TKEtrdecom}\emph{a},\emph{d}) and was envisioned to be composed of mainly VLSMs/superstructures. No expressions exist for wall-normal profiles of the streamwise TI induced by such turbulence, despite the growing research interest in these types of large-scale turbulent motions by the wall-bounded turbulence community. Two decades ago, it was suggested that merging of self-similar LSMs may be one of the mechanisms generating VLSMs and superstructures \citep{adrian:2000a}. Only in recent studies however, it was concluded that wall-attached structures are related to an invariant solution of the Navier-Stokes equations and that those structures comprise families of self-sustaining motions that are consistent with self-similar, wall-attached eddy-scalings \citep{hwang:2016a,cossu:2017a}. Prior to that, a number of studies used linearized Navier-Stokes equations to show that long streaky motions can be amplified at all length scales \citep[][among others]{delalamo:2006a,hwang:2010a,hwang:2010b}, and this is consistent with the recent findings that the large-scale outer structures involve self-sustaining mechanisms \citep{degiovanetti:2017a}. Future studies remain necessary to reveal Reynolds number scalings by way of resolving their spatial and temporal dynamics \citep{kerherve:2017a} and by way of using promising techniques such as variational mode decompositions \citep{wang:2018a}. For now, a full-empirical formulation, judiciously chosen as a parabolic relation with logarithmic argument $\ln(z)$ via (\ref{eq:G}), was found to fit the data.
\begin{eqnarray}
 \label{eq:G}
 \overline{u^2}^+_{\rm G}\left(z^+\right) = G_1 + G_2\ln\left(z^+\right) + G_3\ln\left(z^+\right)^2.
\end{eqnarray}
Figure~\ref{fig:TKEtrdecom}(\emph{b}) presents the five fits of (\ref{eq:G}), for $z_T \leqslant z \leqslant 0.15\delta$. Generally, its energy content increases with $Re_\tau$, but a consistent monotonic trend is absent, owing to the experimental difficulties in acquiring repeatable and converged data at very large wavelengths \citep{samie:2017phd}. Nevertheless, in light of our work, this component is responsible for the secondary peak (or hump) in $\overline{u^2}$ \citep{hultmark:2012a,vallikivi:2015a,willert:2017a,samie:2018a}. \citet{marusic:2013a} observed that a lower bound of the logarithmic region resembled the $z^+ \propto Re_\tau^{\nicefrac{1}{2}}$ dependence \citep{sreenivasan:1997a,wei:2005a,klewicki:2009a,morrill:2017a}, which is in agreement with the peak locations of $\overline{u^2}_{\rm G}$ (see Figure~\ref{fig:KGfit}\emph{b}). This explains that a steeper logarithmic decay---than one with $A_1 = 0.975$---has been observed in raw, non-decomposed $\overline{u^2}$ profiles \citep[$A_1 = 1.26$ in][]{marusic:2013a}. Namely, we here suggest that this steeper decay beyond $z^+ \propto Re_\tau^{\nicefrac{1}{2}}$ included a logarithmic energy decay of global/VLSM-type energy, superimposed on top of the attached-eddy decay. Interestingly, \citet{hwang:2015a} identified two self-similar families of structures, described as the LSMs and VLSMs, from channel flow simulations at relatively low Reynolds numbers up to $Re_\tau \approx 2\,000$. A self-similar structure of the VLSM content (alongside the self-similar structure of LSMs) supports our hypothesis of two dominant logarithmic decays (and recall that those will only be apparent at high $Re_\tau$, when there is a sufficient wall-normal range for a logarithmic range of scales and when component $\overline{u^2}_\mathcal{W}(z)$ tends to a constant via (\ref{eq:K}) in the upper portion of the logarithmic layer).

\section{Concluding remarks}\label{sec:concl}
\begin{figure} 
\vspace{10pt}
\centering
\includegraphics[width = 0.999\textwidth]{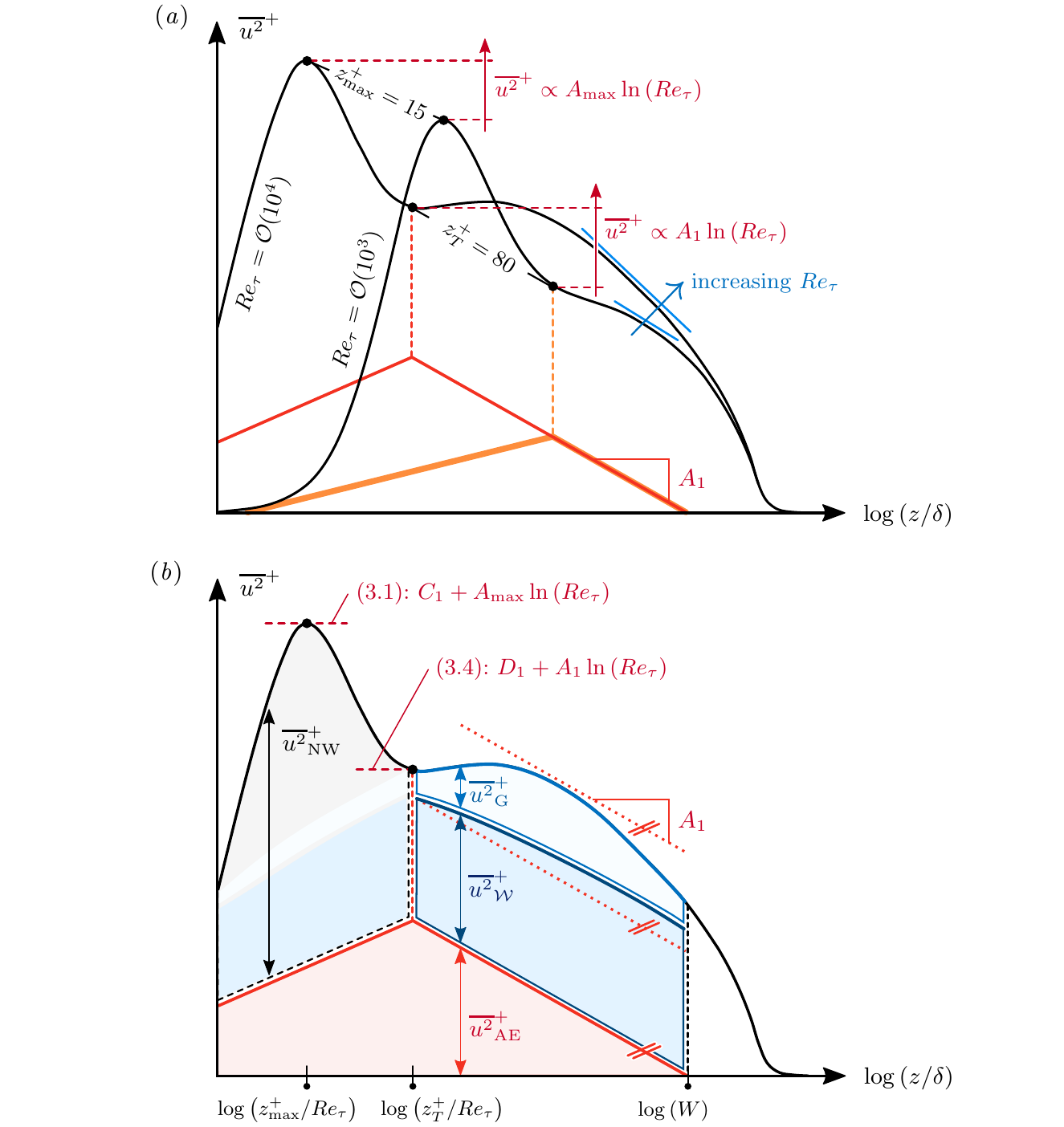}
   \caption{Decomposed structure of $\overline{u^2}(z)$ in ZPG turbulent boundary layers. (\emph{a}) A low and high Reynolds-number turbulent boundary layer profile and (\emph{b}) a breakdown of the streamwise turbulence intensity into three additive contributions in the logarithmic region ($z_T < z < W/\delta$), and two contributions in the near-wall region ($z < z_T$).}
   \label{fig:TKEcon}
\end{figure}
A breakdown of the streamwise TI was assessed through the use of data-driven spectral filters for the streamwise velocity fluctuations $u$ (derived and applied in Part~1). Within the logarithmic region, here taken from $z^+_T = 80$ up to $z/\delta = 0.15$, the streamwise TI from three additive contributions is summarized in Figure~\ref{fig:TKEcon}. The main outcomes of this work are listed as follows.\\[-8pt]
\begin{enumerate}[labelwidth=0.65cm,labelindent=0pt,leftmargin=0.65cm,label=(\roman*),align=left]
\item \noindent Scaling trends of the TI reflecting wall-attached, self-similar eddying motions, revealed evidence for a logarithmic scaling following (\ref{eq:u2logcor}) with $A_1 = 0.975$ (constant over for the range of investigation: $2\,800 \apprle Re_\tau \apprle 19\,300$). A logarithmic decay via $\overline{u^2}_{\rm AE} \propto A_1\ln(z/\delta)$ had to be assumed, because the wall-attached turbulence does comprise a signature of global/VLSM-type energy. It was hypothesized that this energy masks a true logarithmic region in the TI profiles, due to a bulge of $\overline{u^2}_{\rm G}$ energy (Figure~\ref{fig:TKEcon}\emph{b}).\\[-10pt]
\item \noindent Constant $A_1 = 0.975$ is consistent with the growth of the near-wall TI, under the following assumptions: (1) the lower bound of the logarithmic region at which attached-eddy structures become influenced by viscosity scales in inner units (\emph{e.g.} $z^+_T = 80$); (2) below $z^+_T$ the energy-footprint of wall-attached turbulence decays following (\ref{eq:AEslope1}); (3) the near-wall TI-growth with $Re_\tau$ is solely caused by the footprint of the self-similar attached eddies. Assumption (3) implies that the very large-scale outer motions have a negligible contribution to the streamwise TI in the near-wall region (recall \S\,3), at least for the $Re_\tau$ range considered in this study. When accepting these assumptions, the maximum in the near-wall TI profile at $z^+_{\rm max} \approx 15$ disappears as a global maximum at $Re_\tau \approx 3.2 \times 10^7$ following a simple extrapolation (Figure~\ref{fig:TKEnearwall}). Figure~\ref{fig:TKEcon}(\emph{a}) illustrates the attached-eddy scaling in relation to the growth of the near-wall TI. \\[-10pt]
\item \noindent Two components other than the attached-eddy energy are present in the logarithmic region. The stochastically wall-incoherent energy scales following $\overline{u^2}^+_{\mathcal{W}}(z^+) = K_1 - K_2/\sqrt{z^+}$, with $K_1 = 4.01$ and $K_2 = 10.13$. This semi-empirical relation describes Kolmogorov turbulence residing at scales bounded by a $z$-scaled limit and a dissipation limit. When $Re_\tau \rightarrow \infty$, this energy asymptotes to $K_1 = 4.01$ at large $z^+$. A large-scale component $\overline{u^2}_{\rm G}$ comprises global/VLSM-type energy. Other than that this energy seems weakly dependent on $Re_\tau$ (Figure~\ref{fig:KGfit}\emph{b}), definite scaling trends cannot be provided and require future research. \\[-8pt]
\end{enumerate}
Our current work may assist in the development of future data-driven models for the streamwise TI in ZPG TBL flow. Due to dissimilar scalings present over different ranges of the velocity energy spectra of the streamwise velocity $u$, an approach of considering individual sub-components of the streamwise TI, each having their spectral scaling, may be promising for new models. Our current work presented a breakdown of the streamwise TI in the logarithmic region, into three components: a semi-empirical small-scale component comprising Kolmogorov-type turbulence, a model-based component following the AEH, and a remaining contribution of non-self-similar global/VLSM-type energy. When in the near-wall region the footprint of the self-similar attached-eddy contribution is superposed on a universal contribution $\overline{u^2}_{\rm NW}$, the Reynolds number-growth of the near-wall TI can be related to an attached-eddy scaling.

\section*{Acknowledgements}
We gratefully acknowledge the Australian Research Council for financial support and are appreciative of the publicly available DNS data of \citet*{sillero:2013a}. We would also like to give special thanks to Jason Monty, Dominik Krug, Dileep Chandran and Hassan Nagib for helpful discussions on the content of the manuscript.

\bibliographystyle{jfm}
\bibliography{bibtex_database}

\end{document}